\documentclass[twocolumn,prc,amsmath,amssymb,aps,showpacs,superscriptaddress]{revtex4-1}
\usepackage{graphicx}
\usepackage{dcolumn}
\usepackage{bm}
\usepackage{}[multicol]

\newcommand{\eek}{$(e,e^{\prime}K^{+})$}
\newcommand{\pik}{$(\pi^{+},K^{+})$}
\newcommand{\kpi}{$(K^{-},\pi^{-})$}

\newcommand{\cerenkov}{Cherenkov} 
\newcommand{\moller}{M\o ller}
\newcommand{\bl}{$B_{\Lambda}$}

\newcommand*{\TOHOKU}{Department of Physics, Graduate School of Science, Tohoku University, Sendai, Miyagi 980-8578, Japan}
\newcommand*{\TOHOKUMED}{Tohoku Medical and Pharmaceutical University, Sendai, Miyagi 981-8558, Japan}
\newcommand*{\Hampton}{Department of Physics, Hampton University, Hampton, VA 23668, USA}
\newcommand*{\FIU}{Department of Physics, Florida International University, Miami, FL 27411, USA}
\newcommand*{\JLAB}{Thomas Jefferson National Accelerator Facility (JLab), Newport News, VA 23606, USA}
\newcommand*{\KEK}{Accelerator Laboratory, High Energy Accelerator Research Organization (KEK), Tsukuba, Ibaraki 305-0801, Japan}
\newcommand*{\ELPH}{Research Center for Electron Photon Science (ELPH), Tohoku University, Sendai, Miyagi 982-0826, Japan}
\newcommand*{\INES}{Institute of Nuclear Energy Safety, Shushan, Anhui, Hefei 230031, China}
\newcommand*{\GUNMA}{Gunma University Heavy Ion Medical Center,  Maebashi, Gunma 371-8511, Japan}
\newcommand*{\JASRI}{Japan Synchrotron Radiation Research Institute (JASRI), Sayo-gun, Hyogo 679-5198, Japan}

\usepackage{lineno}

\begin{document}
\title{Experimental techniques and performance of \\$\Lambda$-hypernuclear spectroscopy with the {\eek} reaction}
\author{T.~Gogami}\thanks{Corresponding author}\email{gogami@lambda.phys.tohoku.ac.jp}
\address{\TOHOKU}
\author{C.~Chen}\thanks{Present address: {\it \INES}}
\affiliation{\Hampton}
\author{Y.~Fujii}\thanks{Present address: {\it \TOHOKUMED}}
\author{O.~Hashimoto}\thanks{Deceased}
\affiliation{\TOHOKU}
\author{M.~Kaneta}
\affiliation{\TOHOKU}
\author{D.~Kawama}
\affiliation{\TOHOKU}
\author{T.~Maruta} \thanks{Present address: {\it \KEK}}
\affiliation{\TOHOKU}
\author{A.~Matsumura} \thanks{Present address: {\it \GUNMA}}
\affiliation{\TOHOKU}
\author{S.~Nagao}
\affiliation{\TOHOKU}
\author{S.~N.~Nakamura}
\affiliation{\TOHOKU}
\author{Y.~Okayasu} \thanks{Present address: {\it \JASRI}}
\affiliation{\TOHOKU}
\author{J.~Reinhold}
\affiliation{\FIU}
\author{L.~Tang}
\affiliation{\Hampton}
\author{K.~Tsukada} \thanks{Present address: {\it \ELPH}}
\affiliation{\TOHOKU}
\author{S.~A.~Wood} 
\affiliation{\JLAB}
\author{L.~Yuan}
\affiliation{\Hampton}
%
\collaboration{ Detector construction and analysis team of HKS collaboration }
\date{\today}

\begin{abstract}
  The missing-mass spectroscopy of $\Lambda$ hypernuclei
  via the {\eek} reaction has been developed
  through experiments at JLab Halls A and C 
  in the last two decades.
  For the latest experiment, E05-115 in Hall C,
  we developed a new spectrometer system consisting of the HKS and HES;
  resulting in the best energy resolution ($\Delta E \simeq 0.5$-MeV FWHM)
  and $B_{\Lambda}$ accuracy ($\Delta B_{\Lambda} \leq 0.2$~MeV)
  in $\Lambda$-hypernuclear reaction spectroscopy.
  This paper describes the characteristics of the {\eek} reaction compared to
  other reactions and experimental methods.
  In addition, the experimental apparatus,
  some of the important analyses such as the semi-automated calibration of
  absolute energy scale,
  and the performance achieved in E05-115 are presented.
\end{abstract}

\maketitle

\section{Introduction}
Compared to the nucleon-nucleon ($NN$) interaction,
hyperon-nucleon ($YN$) and $YY$ interactions 
are difficult to investigate with free scattering experiments 
due to experimental difficulties 
originating from the short lifetimes of hyperons 
(e.g. $c\tau=7.89$~cm for $\Lambda$).
Therefore, these interactions have been studied primarily via
measurements of
energy levels and transitions of hypernuclei.
Almost 40 species of $\Lambda$ hypernuclei up to 
a mass number of $A=209$ have been measured 
to date~\cite{cite:hashimototamura, cite:feliciello_nagae, cite:gal_hun_mil}
in order to investigate the effective 
$\Lambda N$ potential.
However, more precise and systematic measurements are 
needed to deepen our understanding of the $\Lambda N$
interaction. 
Today, scientists investigate $\Lambda$ hypernuclei
with various types of beams: 
1)~hadron beams at the Japan Proton Accelerator Research Complex
(J-PARC)~\cite{cite:sugimura,cite:yamasan,cite:honda},
2)~heavy-ion beams at GSI
~\cite{cite:hyphi,cite:rappold,cite:rappold2},
3)~heavy-ion colliders at 
the Brookhaven National Laboratory (BNL) 
Relativistic Heavy Ion Collider (RHIC)~\cite{cite:star} 
and the CERN Large Hadron Collider (LHC)~\cite{cite:alice},
and 4)~electron beams at the Mainz Microtron (MAMI)~\cite{cite:patrick, cite:florian, cite:patrick_hyp} and
the Thomas Jefferson National Accelerator Facility
(JLab)~\cite{cite:7LHe,cite:12LB,cite:10LBe,cite:7LHe_2}.
These different reactions are complementary and 
allow us to use their sensitivities 
to study particular nuclear features of interest.

The present paper describes experimental methodology, apparatus and 
some analyses of the latest hypernuclear experiment (Experiment JLab E05-115)
via the {\eek} reaction. 
Section~\ref{sec:role_of_eek} shows the role of missing-mass spectroscopy 
by means of electron scattering compared
to other experimental investigations of $\Lambda$ hypernuclei.
In Sec.~\ref{sec:expset}, the kinematics,
apparatus and setup of JLab E05-115 are described. 
Section~\ref{sec:analysis} shows some of the data analyses
such as $K^{+}$ identification and energy-scale calibration etc.
The achieved missing-mass resolution comparing to design performance
is shown in Sec.~\ref{sec:results}, followed by the conclusion in Sec.~\ref{sec:conclusion}. 


\section{Missing-mass spectroscopy with the ${\bf (e,e^{\prime}K^{+})}$ reaction}
\label{sec:role_of_eek}
$\Lambda$ hypernuclei were first found in nuclear emulsions
that were exposed to cosmic rays~\cite{cite:danysz_p}.
Later, the $\Lambda$ binding energies {\bl},
defined in Eq.~(\ref{eq:binding_energy}), 
of $A\leq15$ hypernuclei were obtained in 
experiments with nuclear emulsions exposed to mesic beams 
such as $K^{-}$~\cite{cite:juric}.
A typical accuracy on the {\bl} determination 
is approximately $\Delta B_{\Lambda} \leq 0.1$~MeV including systematic errors. 
Except for a few cases,
emulsion experiments were able to determine only the 
ground-state {\bl} as they
derived the {\bl} by tracing weak decay processes 
of $\Lambda$ hypernuclei which take a longer time than 
deexcitations emitting $\gamma$-rays and neutrons. 
In addition, with the emulsion technique,
the complexity of decay sequences from heavier 
hypernuclei prevented {\bl} measurements with the mass numbers 
larger than fifteen.

The ground state and excitation energies for light and
heavy hypernuclear systems
up to $A=209$ were investigated by {\kpi} and {\pik} reaction 
spectroscopy using the missing-mass method.
One of novel results is a clear observation of 
shell structures even deeply inside nuclei
for heavy hypernuclear systems~\cite{cite:hasegawa2,cite:hotchi} 
which are not observed by spectroscopy 
of ordinary nuclei due to the large natural widths of the states.
This is due to the fact that a $\Lambda$ can reside
in a deep orbit occupied by nucleons 
since a single embedded $\Lambda$ is not subject 
to the Pauli Principle from nucleons.
The energy resolution in the resulting hypernuclear structures 
was limited to a few MeV FWHM and was
dominated by contributions from the quality of the secondary meson beams. 
Moreover, the energy scales of all {\pik} experiments 
were calibrated to the published result of $^{12}_{\Lambda}$C 
from the emulsion experiments $B_{\Lambda}(^{12}_{\Lambda}{\rm C}) = 10.76\pm0.19$~MeV
which is a mean value of six selected events~\cite{cite:12LC_1,cite:12LC_2}.
In the {\pik} experiments, therefore, 
the error on the reported $B_{\Lambda}(^{12}_{\Lambda}{\rm C})$ contributed
to the {\bl} measurement, and it resulted in 
a $\geq0.5$-MeV systematic error on $B_{\Lambda}$~\cite{cite:hasegawa2}.
It is worth noting that the reported 
$B_{\Lambda}(^{12}_{\Lambda}{\rm C})$ indicated to be
shifted by about 0.54~MeV according to a careful
comparison among results from the emulsion,
{\pik} and {\eek} experiments~\cite{cite:10LBe}.
Recently, totally independent analysis of (K$^{-}_{\rm stop}$,$\pi^{-}$) and
{\pik} data confirmed the existence of the 0.6-MeV difference
between them~\cite{cite:finuda}.
Thus, the {\bl} results from the {\pik} experiments need
a correction of about half MeV.

Recently, hypernuclei of $A=3,4$ were studied in an
experiment with a heavy-ion beam impinging on a fixed target at GSI 
using the invariant-mass technique~\cite{cite:hyphi,cite:rappold,cite:rappold2}.
Also, observations of hypertriton ($^{3}_{\Lambda}$H) 
and anti-hypertriton [$^{3}_{\overline{\Lambda}}{\rm \overline{H}}$ 
($^{2}\overline{{\rm H}}+\overline{\Lambda}$)] nuclei
were reported by the STAR Collaboration at RHIC~\cite{cite:star} 
and the ALICE Collaboration at LHC~\cite{cite:alice} using 
heavy-ion collisions. 
Invariant-mass spectroscopy with heavy-ion beams and 
colliders has the potential to access exotic $\Lambda$ hypernuclei 
far from the nuclear-stable valley.  Such exotic hypernuclei are not
accessible with reaction spectroscopy.
However, the energy resolution and {\bl}
accuracy are larger than 5-MeV FWHM and a few MeV, respectively.

Missing-mass spectroscopy using
an electron beam allows us to achieve 
a better energy resolution ($\simeq 0.5$~MeV FWHM)
and {\bl} accuracy ($\Delta B_{\Lambda} \leq 0.2$~MeV)~\cite{cite:12LB}
than with currently available meson beams.
The properties of the primary electron beam (small emittance and $\Delta E/E$) 
result in a better energy resolution in a hypernuclear spectrum. 
While the production cross section for $\Lambda$ hypernuclei
from the {\eek} reaction is smaller than for the {\kpi}
and {\pik} reactions by 2--3~orders of magnitude~\cite{cite:mot_sot_ito},
this is compensated by the intense primary beam.  Furthermore, the
high intensity allows us to use 
thinner-production targets (order of 0.1~g/cm$^{2}$), contributing
to improvement of the energy resolution.
From the view point of the energy resolution, 
$\gamma$-ray spectroscopy which 
measures deexcitation $\gamma$-rays from $\Lambda$ hypernuclei 
is far much better, with resolutions down to a few keV 
(FWHM)~\cite{cite:yamasan,cite:tamura,cite:tanida,cite:akikawa,cite:ukai,cite:hosomi,cite:yang_tamura}. 
Detailed low-lying structures of $\Lambda$ hypernuclei with $A\leq19$ 
have been investigated with $\gamma$-ray spectroscopy.
However, $\gamma$-ray spectroscopy cannot determine {\bl} 
since it measures only energy spacings.

A proton is converted into a $\Lambda$ in the {\eek} reaction 
while it is a neutron that is converted in {\kpi} and {\pik} reactions.
This feature of the {\eek} reaction enabled us 
to accurately calibrate the energy scale well using 
$\Lambda$ and $\Sigma^{0}$ production from a hydrogen target.
The masses of these calibration references are 
known to be $M(\Lambda) = 1115.683\pm0.006$
and $M(\Sigma^{0}) = 1192.642\pm0.024$~MeV~\cite{cite:pdg}
with errors much smaller 
than that of the reported {\bl}($^{12}_{\Lambda}$C) which 
was, as noted above, used as the {\bl}-measurement 
reference for {\pik} experiments. 
We achieved a total systematic uncertainty on {\bl}
to be $0.11$~MeV (typically $\Delta B_{\Lambda}\leq0.2$~MeV after statistical contribution
are included)
after energy-scale calibration 
in the present experiment (JLab E05-115) as shown in Sec.~\ref{sec:energy_calib}.
On the other hand, in the meson-beam spectroscopy,
such elementary processes cannot be used as the energy-scale reference
because a neutron target does not exist. 
The high-accuracy {\bl} determination 
by the decay $\pi^{-}$ spectroscopy, 
which measures $\pi^{-}$ momenta
from two-body weak decays of $\Lambda$ hypernuclei at rest
for the mass reconstruction, 
has been proven by measuring $^{4}_{\Lambda}$H
at MAMI~\cite{cite:patrick, cite:florian, cite:patrick_hyp}. 
The energy resolution and
{\bl} accuracy achieved were
$\Delta E=0.1$ and $\Delta B_{\Lambda} \leq 0.1$~MeV,
respectively, in decay $\pi^{-}$ spectroscopy.

Furthermore, the {\eek} reaction can investigate 
hypernuclei whose isotopic mirror partners have been well studied 
by meson-beam experiments. 
With a $^{12}$C target, for example, the {\eek}$^{12}_{\Lambda}$B 
and {\pik}$^{12}_{\Lambda}$C hypernuclei can be measured 
and compared with each other.
Such a comparison between mirror hypernuclei provides 
insight into charge symmetry breaking (CSB) in the $\Lambda N$ interaction\cite{cite:gal,cite:dani}.
The HKS Collaboration reported on the results of 
$^{7}_{\Lambda}$He~\cite{cite:7LHe,cite:7LHe_2} 
and $^{10}_{\Lambda}$Be~\cite{cite:10LBe} 
along with discussions of $\Lambda$ hypernuclear CSB 
in the $A=7$ isotriplet ($T=1$) 
($^{7}_{\Lambda}$He, $^{7}_{\Lambda}$Li$^{*}$, $^{7}_{\Lambda}$Be)
and $A=10$, $T=1/2$ ($^{10}_{\Lambda}$Be, $^{10}_{\Lambda}$B) systems, 
respectively. 
\begin{table*}[!htbp]
  \begin{center}
    \caption{Typical energy resolutions ($\Delta E$),
      $B_{\Lambda}$ accuracy ($\Delta B_{\Lambda}$),
      and mass numbers ($A$) measured
      in the various hypernuclear experiments shown in the text.}
    \label{tab:eek_others_comp}
    \begin{tabular}{|c|c|c|c|c|}
      \hline
      \multicolumn{2}{|c|}{Experimental} & $\Delta E$& $\Delta B_{\Lambda}$ & $A$\\ 
      \multicolumn{2}{|c|}{technique} & (FWHM)~(keV)      & (keV)                  & (so far)\\ \hline \hline
      \multicolumn{2}{|c|}{Emulsion}   & - & $\leq 100$ & $\leq 15$  \\ \hline 
      Missing-mass & {\kpi}, {\pik} & $\geq 1000$ & $\leq 1000$ & $\leq 209$ \\ \cline{2-5}
      spectroscopy    & {\eek} & $\simeq 500$ & $\leq 200$ & $\leq 52$ \\ \hline 
      \multicolumn{2}{|c|}{Invariant-mass spectroscopy} & $> 5000$ & a few 1000 & $\leq 4$ \\ 
      \multicolumn{2}{|c|}{with heavy-ion beam/collider}  & & & \\ \hline
      \multicolumn{2}{|c|}{$\gamma$-ray spectroscopy} & a few & - & $\leq 19$ \\ \hline 
      \multicolumn{2}{|c|}{Decay $\pi^{-}$ spectroscopy} & $\simeq 100$ & $\leq 100$ & 4 \\ 
      \hline 
    \end{tabular}
  \end{center}
\end{table*}

Typical energy resolutions, $B_{\Lambda}$ accuracy, 
and mass numbers of hypernuclei measured in the various hypernuclear experiments
described above are tabulated in Table~\ref{tab:eek_others_comp}.
For $A > 15$, experiments using reactions other than {\eek}
have not measured $\Lambda$ binding energies with accuracy or with energy
resolutions much better than one MeV.
Improving the accuracy and resolution provides insight into  
1)~the many-body baryon interactions which are expected to 
act an important role particularly in high density 
nuclear matters such as neutron stars~\cite{cite:yamamoto_neustar}, 
2)~the dynamics of nuclear deformation by 
adding a $\Lambda$ as an impurity~\cite{cite:isaka_def,cite:win,cite:bnlu,cite:xue},
3)~the $p$-shell hypernuclear 
CSB~\cite{cite:gal,cite:hiyama1,cite:hiyama_10LB},
and so on.
In addition, sub-MeV resolution is necessary to 
resolve particular $\Lambda$-hypernuclear structures
that are
due to effects such as a core-configuration mixing and spin-orbit 
splitting~\cite{cite:bydzovsky}.
In terms of required momentum resolution and acceptance
of a magnetic-spectrometer system in addition to 
the beam quality, 
JLab is a unique facility, at the moment, 
to perform spectroscopic studies for 
medium to heavy $\Lambda$ hypernuclei 
with sub-MeV energy resolution and 
{\bl} accuracy of a few hundred keV or better~\cite{cite:kusaka-fujita,cite:loi208}.
The JLab Hall C facility has, so far, measured hypernuclei
$^{7}_{\Lambda}$He~\cite{cite:7LHe,cite:7LHe_2}, 
$^{9}_{\Lambda}$Li~\cite{cite:toshi}, 
$^{10}_{\Lambda}$Be~\cite{cite:10LBe},
$^{12}_{\Lambda}$B~\cite{cite:miyoshi,cite:lulin,cite:12LB},
$^{28}_{\Lambda}$Al~\cite{cite:28LAl}, 
and $^{52}_{\Lambda}$V~\cite{cite:52LV}.
In addition, in Hall A at JLab, which covers 
different kinematical region than Hall C, 
$^{9}_{\Lambda}$Li~\cite{cite:9LLi}, 
$^{12}_{\Lambda}$B~\cite{cite:iodice}, 
and $^{16}_{\Lambda}$N~\cite{cite:cusanno} were measured.


\section{Experimental setup and apparatus}
\label{sec:expset}
\subsection{Kinematics}
\begin{figure}[!htbp]
  \begin{center}
    \includegraphics[width=8.0cm]{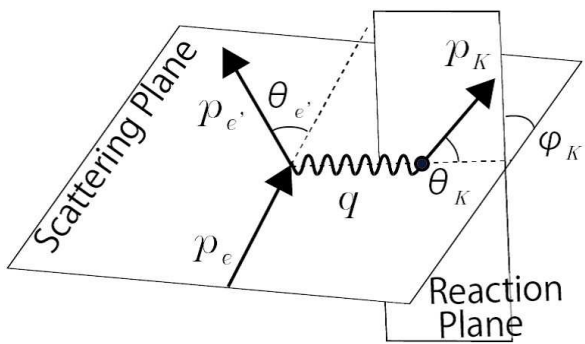}
    \caption{A schematic of the {\eek} reaction.}
    \label{fig:eek}
  \end{center}
\end{figure}

Electroproduction is related to photoproduction 
through a virtual photon produced in the ($e,e^{\prime}$) 
reaction~\cite{cite:sotona,cite:hunger,cite:xu}.
Figure~\ref{fig:eek} shows a schematic of the {\eek} reaction. 
The $p$ shown in the figure denotes the four momentum of each particle, 
and $q=p_{e}-p_{e^{\prime}}$ is the four-momentum transfer to the virtual photon. 
The energy and momentum of the virtual photon are defined as:
\begin{eqnarray}
  \omega &=& E_{e} - E_{e^{\prime}} \hspace{0.1cm},\\
  \vec{q} &=& \vec{p}_{e} - \vec{p}_{e^{\prime}} \hspace{0.2cm}.
\end{eqnarray}
The triple-differential cross section for $\Lambda$ hypernuclear 
production is described by the following form~\cite{cite:sotona, cite:hunger}:
\begin{eqnarray}
  \label{eq:cs_eek}
  \frac{d^{3}\sigma}{dE_{e^{\prime}}d\Omega_{e^{\prime}}d\Omega_{K}} = 
  \Gamma \Bigl( \frac{d\sigma_{U}}{d\Omega_{K}} 
  + \epsilon_{L} \frac{d\sigma_{L}}{d\Omega_{K}}
  + \epsilon \frac{d\sigma_{P}}{d\Omega_{K}} \nonumber \\ 
  + \sqrt{\epsilon_{L}(1+\epsilon)} \frac{d\sigma_{I}}{d\Omega_{K}} \Bigr)
\end{eqnarray}
where $\sigma_{U}$, $\sigma_{L}$, $\sigma_{P}$ and $\sigma_{I}$ are the 
unpolarized transverse, longitudinal, polarized transverse and interference 
cross sections, respectively. 
The $\Gamma$ is the virtual photon flux represented by: 
\begin{eqnarray}
  \label{eq:vpflux}
  \Gamma = \frac{\alpha}{2\pi^{2}Q^{2}} \frac{E_{\gamma}}{1-\epsilon} \frac{E_{e^{\prime}}}{E_{e}}
\end{eqnarray}
where $\alpha = \frac{e^{2}}{4\pi} = \frac{1}{137}$ and $Q^{2}=-q^{2}>0$.
The virtual photon transverse polarization ($\epsilon$), 
longitudinal polarization ($\epsilon_{L}$), 
and the effective photon energy ($E_{\gamma}$) in 
Eq.~(\ref{eq:cs_eek}) and Eq.~(\ref{eq:vpflux}) are defined as follows:
\begin{eqnarray}
  \epsilon &=& \Bigl( 1 + \frac{2|\vec{q}|^{2}}{Q^{2}} \tan^{2}{\frac{\theta_{e^{\prime}}}{2}}\Bigr)^{-1}, \\
  \epsilon_{L} &=& \frac{Q^{2}}{\omega^{2}}\epsilon, \\
  E_{\gamma} &=& \omega + \frac{q^{2}}{2m_{p}},
\end{eqnarray}
where $\theta_{e^{\prime}}$ is the electron scattering angle in the 
laboratory frame. 
In the case of real photons, only the unpolarized transverse term 
is nonvanishing because $Q^{2}\rightarrow 0$.
In the experimental geometry for JLab E05-115, 
the virtual photon can be treated as almost real as
$Q^{2}$ was quite small 
[$Q^{2} \simeq$ 0.01~(GeV/$c$)$^{2}$, $\epsilon \simeq 0.63$]~\cite{cite:12LB}.

An electron beam 
with an energy of $E_{e}=2.344$~GeV, provided by 
the Continuous Electron Beam Accelerator Facility (CEBAF)
at JLab, was used for the experiment.
In order to maximize the yield of $\Lambda$ hypernuclei, 
the virtual photon energy at $\omega=1.5$~GeV [$\sqrt{s} = 1.92$~GeV for $p$($\gamma^{*},K^{+}$)]
was chosen where the production-cross sections of both
$\Lambda$ and $\Sigma^{0}$ hyperons by photoproduction are large~\cite{cite:bradford}.
Hence, the central momentum of the scattered electron is 
designed to be at $|\vec{p_{e^{\prime}}}| \simeq E_{e^{\prime}}=E_{e}-\omega=0.844$~GeV/$c$.
In this case, the $K^{+}$ momentum is approximately 1.2~GeV/$c$.
%

\begin{figure}[!htbp]
  \begin{center}
    \includegraphics[width=8.6cm]{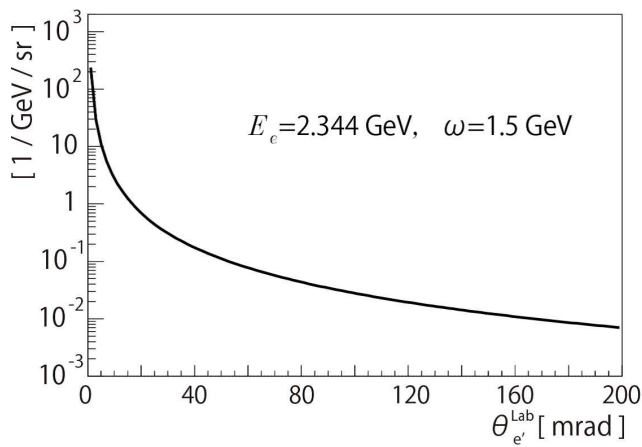}
    \caption{A calculated virtual-photon flux defined in Eq.~(\ref{eq:vpflux}) as a
      function of a scattered-electron angle $\theta_{e^{\prime}}$ in the
      laboratory frame at $E_{e}=2.344$~GeV and $\omega=1.5$~GeV.}
    \label{fig:vpflux}
  \end{center}
\end{figure}
Figure~\ref{fig:vpflux} shows the virtual photon flux 
defined in Eq.~(\ref{eq:vpflux}) as a
function of the scattered angle of $e^{\prime}$ in the 
laboratory frame at $E_{e}=2.344$~GeV and $\omega=1.5$~GeV.
The virtual photon flux is large at the small scattering 
angle of $e^{\prime}$.
At the same time, the small $K^{+}$ scattering angle
yields a large production-cross section for $\Lambda$ hypernuclei~\cite{cite:sotona}. 
Consequently, the detectable scattering angles for both $e^{\prime}$ and $K^{+}$ 
should be as small as possible to maximize the yield of $\Lambda$ hypernuclei. 
For this purpose, a charge separation dipole magnet 
[splitter magnet (SPL)] was installed right after 
the production target to bend the
$K^{+}$ and $e^{\prime}$ in opposite directions
towards each of the magnetic spectrometers
as shown in Sec.~\ref{sec:magnetic_spec}.

\subsection{Magnetic spectrometers}
\label{sec:magnetic_spec}
\begin{figure*}[!htbp]
  \begin{center}
    \includegraphics[width=13.0cm]{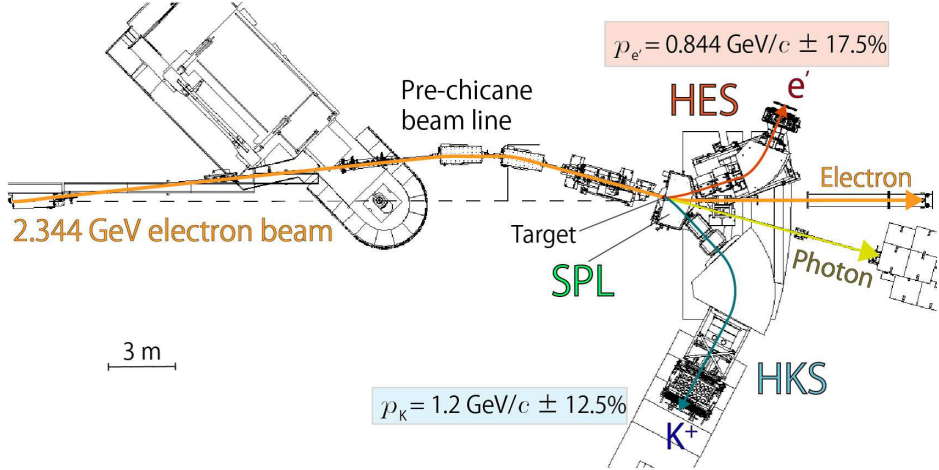}
    \caption{A schematic of the experimental setup of JLab E05-115, 
      which was performed at JLab Hall C in 2009.
      The $E_{e}=2.344$-GeV continuous-wave electron beam, incident
      on the production target located at the SPL entrance, produced
      lambda hypernuclei via the {\eek} reaction.
      $K^{+}$s and scattered electrons at 
      approximately 1~GeV/$c$ were measured and momentum-analyzed 
      in HKS and HES, respectively.
    }
    \label{fig:exp_set}
  \end{center}
\end{figure*}

Figure~\ref{fig:exp_set} shows a schematic of 
the experimental setup of JLab E05-115 in the experimental Hall C.
The electron beam at $E_{e}=2.344$~GeV was incident on the
production target which was installed at the entrance of the SPL.
A $K^{+}$ and scattered electron via the {\eek} reaction 
were bent in opposite directions by the SPL
and were measured with a high-resolution kaon spectrometer 
(HKS)~\cite{cite:gogami, cite:fujii} and a high-resolution electron spectrometer 
(HES), respectively. 
A ``pre''-chicane beam line was designed and used instead of 
the existing beam line at JLab Hall C.
A combination of the pre-chicane beam line
and SPL allowed us to transport unused beams and 
Bremsstrahlung photons generated in the target toward 
beam and photon dumps, respectively, without 
any additional bending magnets between the target and dumps.
On the other hand,
in the previous {\eek} experiment JLab E01-011~\cite{cite:7LHe,cite:12LB}, 
a ``post''-chicane was adopted to transport 
the unused beam to the beam dump.
Though the post-chicane has the merit that one beam dump accepts both
unused electrons and Bremsstrahlung photons,
background particles are likely to be produced by beam halo
which originates from beam broadening in the target. 
Therefore, a larger system of magnet and beam pipe
is necessary for the post-chicane configuration
to suppress the background rate.
The pre-chicane option requires
careful adjustment of electron-beam direction before the target, but
it handles the clean primary beam and the system is compact.
Therefore, a pre-chicane 
was employed for the beam transport in JLab E05-115.

The HKS was constructed and used for the previous 
$\Lambda$ hypernuclear experiment at JLab Hall-C (JLab E01-011), 
and was used again in the present experiment for the $K^{+}$ detection.
The magnet configuration of the HKS was 
two quadrupole and one dipole magnets (Q-Q-D configuration).
Particle detectors which are described in Sec.~\ref{sec:pardet}
were installed downstream of the dipole magnet.
The HES was newly constructed for the present experiment.
The magnet configuration of HES was, like the HKS, Q-Q-D
and the particle detectors were installed behind the dipole magnet. 
SPL was also newly designed and constructed for the present experiment 
and optical matched to the HKS and HES.
The major magnet parameters of the SPL, HKS and HES are summarized in
Table~\ref{tab:hes-magnet}.
\begin{table*}[!htbp]
  \begin{center}
    \caption{Major parameters of magnets of SPL, HKS and HES in the JLab E05-115 experiment.}
    \label{tab:hes-magnet}
    \begin{tabular}{|cc||c|c|c|c|c|c|c|}
      \hline \hline
          & &Magnet & Max. & Max.  & Gap  & Max. & Bore  & Pole  \\ 
          & &weight & current & field  & height & field grad.& radius & length \\
          & &(ton)  &  (A)    & (T) & (mm) & (T/m) & (mm)  & (mm) \\  \hline
      SPL & (D)& 31.7 & 1020& 1.74  & 190 & - & - & -\\ \hline
      HKS & Q1 & 8.2 & 875 & - & - & 6.6 & 120 & 840 \\
          & Q2 & 10.5& 450 & - & - & 4.2 & 145 & 600 \\
          & D  & 210 & 1140& 1.53 & 200 &- & - & 3254 \\ \hline
      HES & Q1 & 2.8 & 800 & - & - & 7.8 & 100 & 600 \\
          & Q2 & 3.1 & 800 & - & - & 5.0 & 125 & 500 \\
          & D  & 36.4& 1065& 1.65  & 194 & - & -  & 2049\\
      \hline \hline
    \end{tabular}
  \end{center}
\end{table*}

One of important features in the present experiment 
is a high-momentum resolution of $\Delta p/p \simeq 2 \times 10^{-4}$
(FWHM) for both $K^{+}$ and $e^{\prime}$ at about 1~GeV/$c$,
owing to optical systems of SPL $+$ HKS and SPL $+$ HES, respectively.
This resulted in an energy resolution of about 0.5~MeV (FWHM)
in the measured hypernuclear structures~\cite{cite:12LB,cite:10LBe}.
Table~\ref{tab:spec_spec} shows 
some of specifications of the spectrometers.
\begin{table*}[!htbp]
  \caption{Key specifications of our spectrometers in JLab E05-115 experiment.}
  \label{tab:spec_spec}
  \begin{tabular}{|c|cc|}
    \hline \hline
    Spectrometer          & SPL+HKS ($K^{+}$) & SPL+HES ($e^{\prime}$) \\ \hline
    Central momentum (GeV/$c$) & 1.200 & 0.844 \\
    Momentum bite & $\pm 12.5\%$ & $\pm 17.5\%$ \\
    Momentum resolution ($\Delta p / p$)  & \multicolumn{2}{c|}{$2 \times 10^{-4}$ (FWHM)} \\
    Angular acceptance in laboratory frame (deg)  & 1--13 & 2--13 \\
    Solid angle at the central momentum (msr) & 8.5   & 7.0 \\
    \hline \hline
  \end{tabular}
\end{table*}

\subsection{Particle Detectors}
\label{sec:pardet}
\begin{figure}[!htbp]
  \begin{center}
    \includegraphics[width=8.6cm]{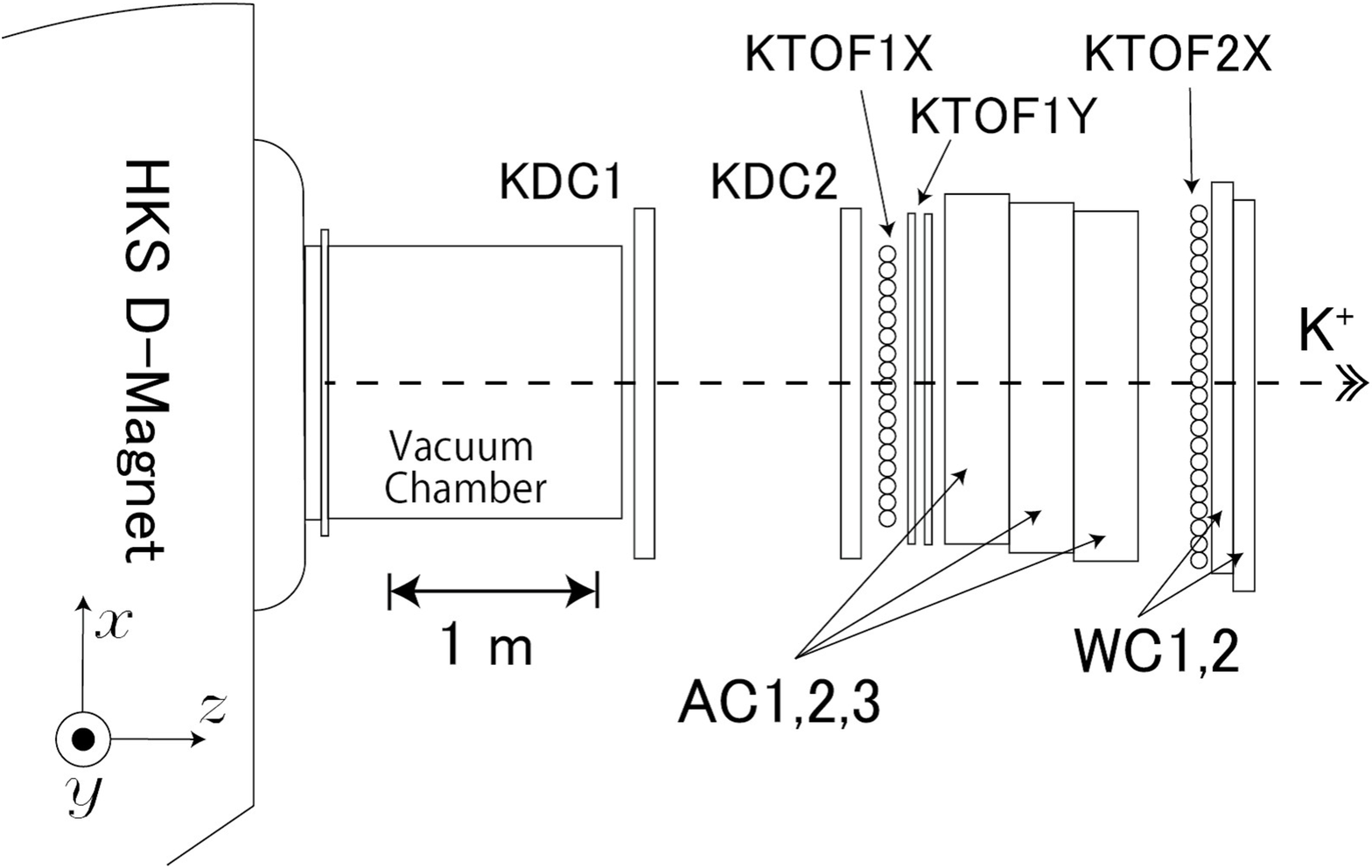}
    \caption{A schematic drawing of the HKS detector system.
    The HKS detector system is composed of 
    two planar-drift chambers (KDC1,2) for a particle tracking, 
    three layers of TOF detectors (KTOF1X, KTOF1Y, KTOF2X) 
    for the data-taking trigger and off-line PID, 
    and two types of {\cerenkov} detectors (AC1--3, WC1,2)
    for both on-line and 
    off-line PID.}
    \label{fig:hks_det}
  \end{center}
\end{figure}

The HKS ($K^{+}$ spectrometer) detector system was composed of 
two drift chambers (KDC1, KDC2) for a particle tracking, 
three layers of time-of-flight (TOF) detectors 
(KTOF1X, KTOF1Y, KTOF2X) used for 
the data-taking trigger and off-line particle identification (PID), 
and two types of {\cerenkov} detectors 
with radiation media of aerogel (refractive index of $n=1.05$)
and water ($n=1.33$) (AC1--3, WC1,2) for both on-line and off-line PID.
Figure~\ref{fig:hks_det} shows a schematic of the HKS detector system,
in which $x$, $y$ and $z$-coordinates in HKS are defined.
KDC1 and KDC2 are identical planar-drift chambers with a cell size of 5~mm.
Each KDC consists of six layers with  
wire configurations of $uu^{\prime}(-60^{\circ})xx^{\prime}(0^{\circ})vv^{\prime}(+60^{\circ})$. 
The primes ( $^{\prime}$ ) denote planes with wires having a half-cell offset, 
and were used to solve the left-right ambiguity in tracking analysis.
Information on position and angle of a particle at
a reference plane, which is defined as a mid-plane between KDC1 and KDC2,
was obtained by the tracking and used for momentum analysis as shown
in Sec.~\ref{sec:mm_recon}.
A typical KDC plane resolution was $\sigma\simeq280$~$\mu$m.
KTOF1X, KTOF1Y and KTOF2X are plastic scintillation detectors with a thickness of 20~mm 
in $z$-direction.
KTOF1X and KTOF2X are segmented by respectively seventeen and eighteen in $x$-direction, 
and KTOF1Y is segmented by nine in $y$-direction, taking 
into account the counting rate in each segment. 
The timing resolutions of KTOF1X, KTOF2X and KTOF1Y were 
obtained to be $\sigma \simeq 70$, $60$, and $110$~ps, respectively, in cosmic-ray tests.

The primary background particles in the HKS were $\pi^{+}$s and protons.
Yields of $\pi^{+}$s and protons were approximately 80:1 and 30:1, respectively, 
relative to $K^{+}$s, when we used an unbiased trigger (CP$_{{\rm trigger}}$ shown in Sec.~\ref{sec:trig}), 
for a 0.451-g/cm$^{2}$ polyethylene target. 
For the desired $\Lambda$ hypernucleus production rate, these
background fractions were too high for our
data acquisition (DAQ) system.
Thus, these background particles needed to be suppressed at the trigger level (on-line).
In order to suppress $\pi^{+}$s and protons on-line, 
we employed three layers of aerogel {\cerenkov} detectors and 
two layers of water {\cerenkov} detectors, respectively.
On-line rejection capabilities for $\pi^{+}$ and proton were 
$5.4\times10^{-3}$ and $1.2\times10^{-1}$, respectively, 
while maintaining a $K^{+}$ survival ratio of $92\%$
in the case of the polyethylene-target data.
For off-line PID, light-yield information of the {\cerenkov} detectors was used 
in addition to reconstructed particle-mass squares
which was obtained by TOF and momentum analyses
as described in Sec.~\ref{kidana}.
Details about the analyses using the {\cerenkov} detectors 
can be found in Ref.~\cite{cite:gogami}.

\begin{figure}[!htbp]
  \begin{center}
    \includegraphics[width=8.6cm]{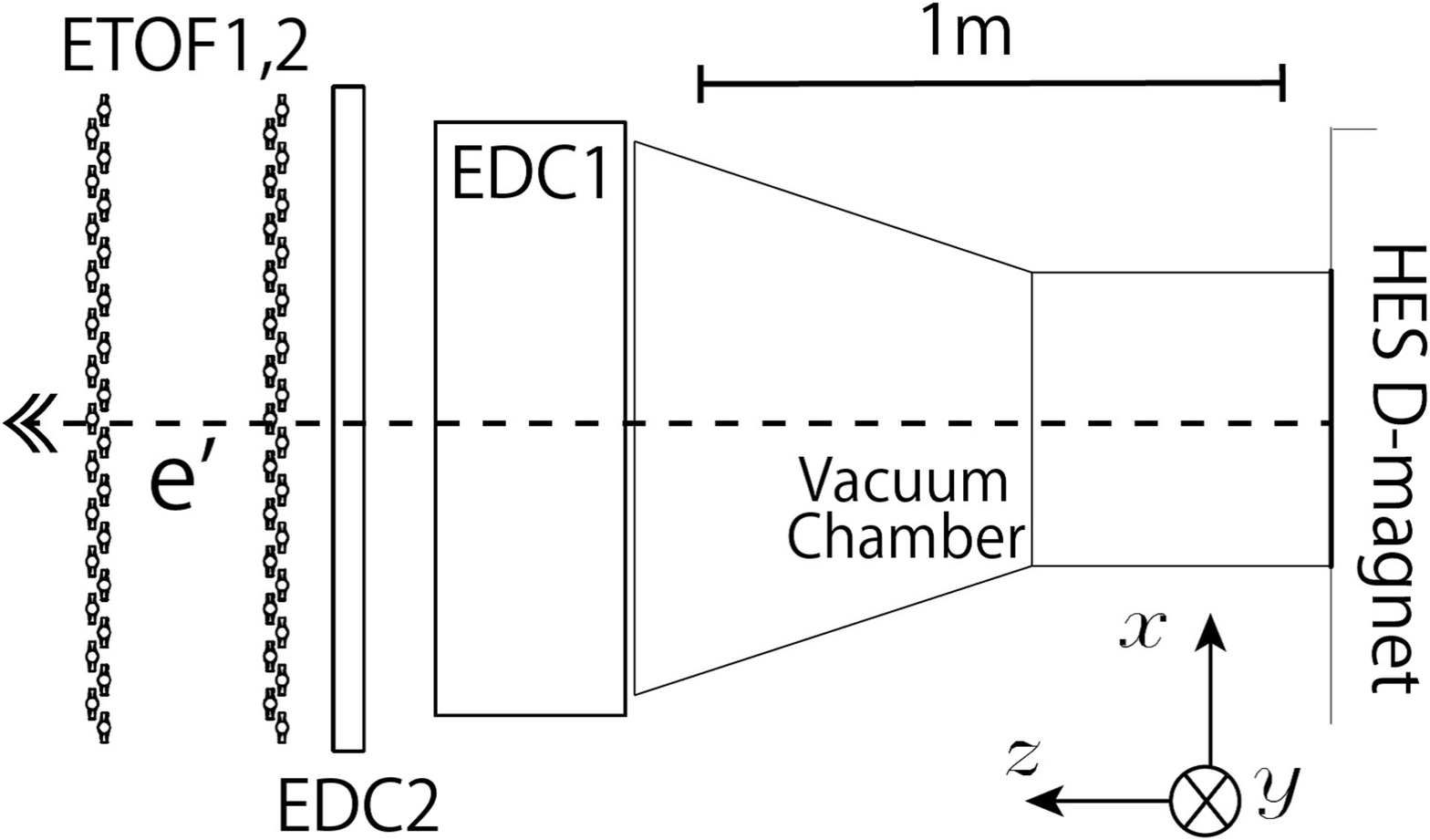}
    \caption{Schematic of the HES detector system.
      The HES detector system is composed of two drift chambers 
      (EDC1,2) for the particle tracking, and two layers of 
      TOF detectors (ETOF1,2) for the data-taking trigger.}
    \label{fig:hes_det}
  \end{center}
\end{figure}
The HES ($e^{\prime}$ spectrometer) detector system consists of 
two drift chambers (EDC1, EDC2) for the particle tracking, and 
two layers of TOF detectors (ETOF1, ETOF2) for the data-taking trigger, 
as shown in Fig.~\ref{fig:hes_det}.
EDC1 is a honeycomb-cell drift chamber with a cell size of 5~mm.
EDC1 consists of ten layers with wire configurations of 
$xx^{\prime}(0^{\circ})uu^{\prime}(-30^{\circ})xx^{\prime}(0^{\circ})vv^{\prime}(+30^{\circ})xx^{\prime}(0^{\circ})$.
The typical plane resolution of EDC1 is approximately $\sigma=220$~$\mu$m.
The HES-reference plane, on which information of position and angle of
particles were used for the momentum analysis in HES,
was defined as the mid-plane of EDC1.
EDC2 is a planar-drift chamber identical to the KDC.
ETOF1 and ETOF2 are plastic scintillation detectors 
each with a thickness of 10~mm in $z$-direction.
The configurations of ETOF1 and ETOF2 are identical. 
Each ETOF is segmented by 29 in $x$-direction, 
taking into account a counting rate in each segment. 
The timing resolution of ETOF 
was obtained to be $\sigma \simeq 100$~ps in cosmic-ray tests.

\subsection{The tilt method in HES}
\label{sec:tilt_method}
The HES detector system was expected to suffer from 
huge amount of background electrons which originate
from electromagnetic processes.
Major sources of background electrons 
were expected to come from 
1)~beam electrons which lose their energies via Bremsstrahlung
process~\cite{cite:tsai}, and 2)~\moller\ scattering 
(elastic electron-electron scattering)~\cite{cite:moller} in the target. 
The reaction cross-sections of these background 
processes are larger at the smaller scattering angle of $e^{\prime}$. 
On the other hand, 
the virtual photon flux, 
which directly relates to the yield of $\Lambda$ 
hypernuclei, is also larger at the small $e^{\prime}$ scattering angle, 
as shown in Fig.~\ref{fig:vpflux}.
Therefore, we attempted to optimize 
the angular acceptance of HES, taking into account the
$S/N$ and yield of $\Lambda$ hypernuclei.
For the purpose, 
we adopted the ``tilt method'', 
which was developed and proven to work sufficiently in the 
previous {\eek} experiment (JLab E01-011)~\cite{cite:miyoshi,cite:lulin}. 

\begin{figure}[!htbp]
  \begin{center}
    \includegraphics[width=8.6cm]{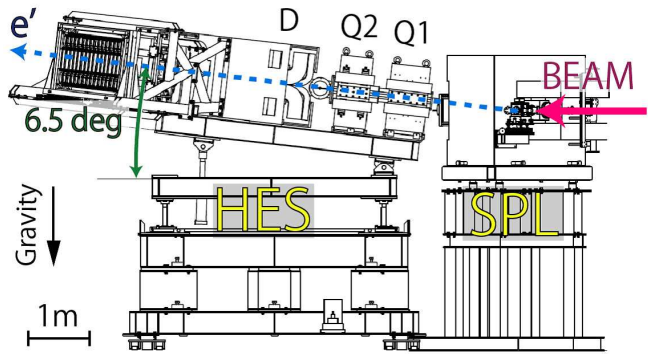}
    \caption{A schematic of the HES side view.
      HES was vertically tilted to avoid 
      the very small $e^{\prime}$ scattering angle, where the $S/N$ is poor.
      The tilt angle at 6.5~degrees was chosen by
      Monte Carlo simulations taking into account
      both yield and $S/N$. }
    \label{fig:hes_tilt}
  \end{center}
\end{figure} 
The tilt method is a method of 
angular acceptance optimization in which
 the magnetic spectrometer is tilted vertically, as shown in Fig.~\ref{fig:hes_tilt}. 
A Monte Carlo simulation was performed to optimize the tilt angle.
A figure-of-merit (FoM) that was used 
for the optimization as a reference was defined as follows: 
\begin{eqnarray}
  \label{eq:hes_fom}
  {\rm FoM} = \frac{R_{{\rm VP}}}{\sqrt{ R_{{\rm Brems}} + R_{\textrm{\moller}} }}, 
\end{eqnarray}
where $R_{{\rm VP}, {\rm Brems}, \textrm{\moller}}$ are counting rates of 
electrons associated with the virtual photon, Bremsstrahlung and \moller\ scattering
in HES. 
\begin{figure}[!htbp]
  \begin{center}
    \includegraphics[width=8.6cm]{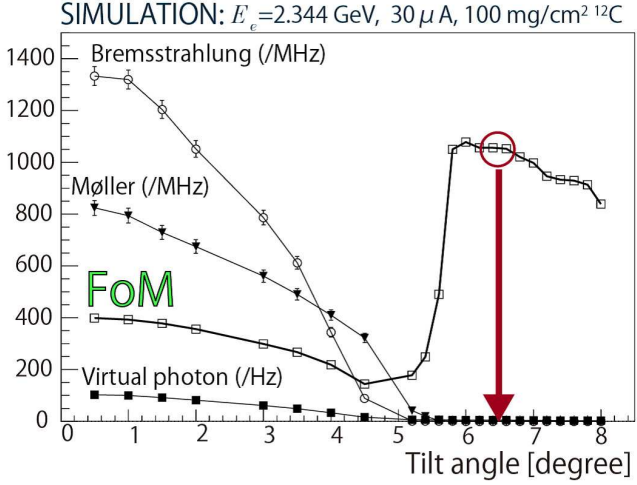}
    \caption{Expected counting rates of electrons 
      associated with the virtual photon ($R_{{\rm VP}}$),
      Bremsstrahlung ($R_{{\rm Brems}}$) and 
      \moller\ scattering ($R_{\textrm{\moller}}$) in HES 
      by a Monte Carlo simulation.
      This simulation assumed a 30-$\mu$A electron beam 
      on a 100-mg/cm$^{2}$ $^{12}$C target at $E_{e}=2.344$~GeV.
      A figure of merit (FoM) defined in Eq.~(\ref{eq:hes_fom}) is also shown,
      and the tilt angle was chosen to be 6.5~degrees. }
    \label{fig:hes_fom_result}
  \end{center}
\end{figure}
Figure~\ref{fig:hes_fom_result} shows the result of the Monte Carlo simulation 
of a 30-$\mu$A electron beam
on a 100-mg/cm$^{2}$ $^{12}$C target at $E_{e}=2.344$~GeV. 
The HES tilt angle was determined to be 6.5~degrees by using this result.
The virtual photon flux [Eq.~(\ref{eq:vpflux})] 
integrated over the acceptance for scattered electrons (Fig.~\ref{fig:sahes})
was evaluated by the Monte Carlo simulation, and  
was found to be $\Gamma^{int} = (5.67\pm0.04)\times10^{-5}$ (/electron) 
for a momentum range of $p_{e^{\prime}}=0.80$--$0.98$~GeV/$c$.

Table~\ref{tab:sn_comp} shows
typical values of beam intensity $I_{b}$, luminosity $L$,
typical angle for scattered electrons $\theta_{e^{\prime}}$,
integrated virtual photon flux $\Gamma ^{int}$, solid-angle acceptance at the central $K^{+}$ momentum
d$\Omega_{K}$, total detection efficiency $\epsilon_{tot}$,
counting rates in $e^{\prime}$ spectrometer $R_{e^{\prime}}$, signal yield per hour per 100~nb/sr,
$S/N$ for the ground-state doublet peak of $^{12}_{\Lambda}$B at the peak position, 
comparing between
E89-009 (without tilt method)~\cite{cite:miyoshi,cite:lulin,cite:miyoshi_dthesis} and
E05-115 (with tilt method)~\cite{cite:12LB, cite:toshi}.
Because of the tilt method,
we were able to increase the luminosity by a factor of 230, while
reducing the counting rate in the scattered electron spectrometer by a factor of 1/100.
Consequently, although the virtual photon flux is smaller by a factor of 0.14
due to the larger $\theta_{e^{\prime}}$,
the yield per a unit time and $S/N$ improved by factors of
60 and 2.5, respectively.
\begin{table*}
  \begin{center}
    \caption{Comparison of experimental conditions in E89-009 (without tilt method)
      and E05-115 (with tilt method).}
    \label{tab:sn_comp}
    \begin{tabular}{|c|c|c|c|c|c|c|c|c|c|}
      \hline \hline
      Experiment & $I_{b}$  & $L$  & $\theta_{e^{\prime}}$  & $\Gamma ^{int}$ & $d\Omega_{K}$  & $\epsilon_{\rm tot}$ & $R_{e^{\prime}}$ & Yield  & $S/N$ \\
      (year) & ($\mu$A) & (cm$^{-2}$s$^{-1}$) & (deg) & (/electron) & (mrad) & & (MHz) & per hour & \\
             &          &                     &       & & & & & per 100~nb/sr& \\ \hline
      E89-009 (2000) & 0.6 & $4.1\times10^{33}$ & 0--4 & $4.0 \times 10^{-4}$ & 5.0 & 0.18 & 200 & 0.5 & 1.1 \\
      E05-115 (2009) & 35  & $9.6\times10^{35}$ & 2--13 & $5.7 \times 10^{-5}$ & 8.5 & 0.17 & 2& 30 & 2.8 \\
      \hline \hline
    \end{tabular}
  \end{center}
\end{table*}

\subsection{Spectrometer acceptance}
\begin{figure}[!htbp]
  \begin{center}
    \includegraphics[width=8.5cm]{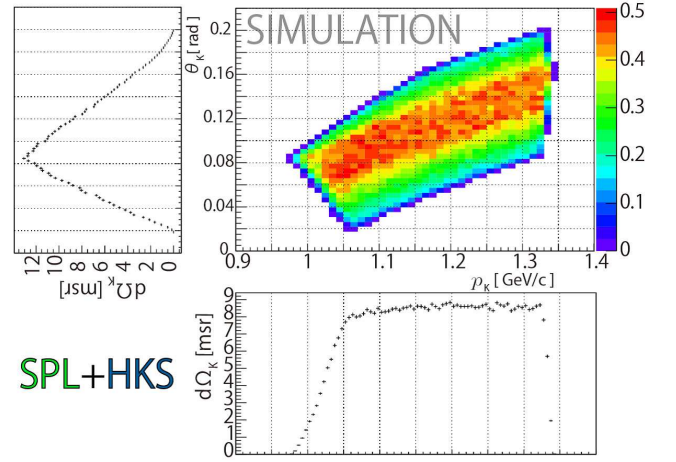}
    \caption{The estimated solid-angle acceptance of SPL $+$ HKS.}
    \label{fig:sahks}
  \end{center}
\end{figure}

\begin{figure}[!htbp]
  \begin{center}
    \includegraphics[width=8.5cm]{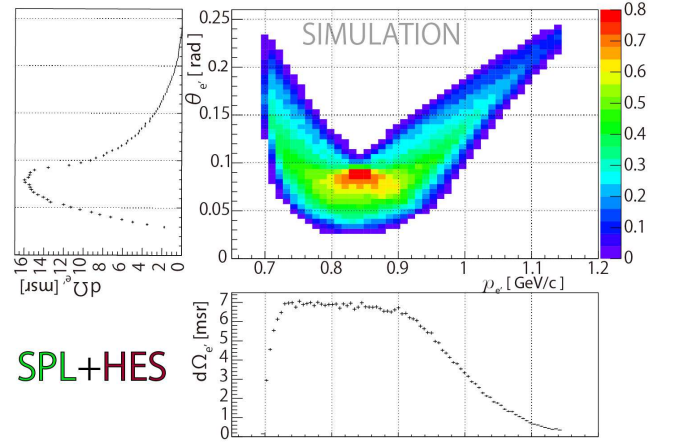}
    \caption{The estimated solid-angle acceptance of SPL $+$ HES.}
    \label{fig:sahes}
  \end{center}
\end{figure}

The acceptance for each SPL $+$ HKS and SPL $+$ HES 
optical system was estimated by the Monte Carlo simulation.
In the simulation, realistic experimental geometries
and magnetic field maps calculated by
Opera3D (TOSCA)~\cite{cite:tosca}
were used.
Figures~\ref{fig:sahks} and \ref{fig:sahes} show 
the estimated solid-angle acceptances for SPL $+$ HKS and SPL $+$ HES, 
respectively.  
The solid-angle acceptances of SPL $+$ HKS and SPL $+$ HES at 
each central momentum are approximately 8.5 and 7.0~msr.

\subsection{Production Target}
\label{sec:target}
In the experiment, a natural carbon target and
isotopically-enriched solid-targets of $^{7}$Li, $^{9}$Be, $^{10}$B, and $^{52}$Cr 
were used for $\Lambda$ hypernuclear production.
In addition, we used polyethylene (CH$_{2}$) and water (H$_{2}$O)
targets to measure $\Lambda$ and $\Sigma^{0}$ production
from hydrogen nuclei.  These targets were used for 
the energy-scale calibration described in Sec.~\ref{sec:energy_calib}.
The targets used in the experiment are summarized in Table~\ref{tab:targets}.
\begin{table*}[!htbp]
  \begin{center}
    \caption{A list of targets used for the JLab E05-115 experiment.}
    \label{tab:targets}
    \begin{tabular}{|c|c|c|c|c|c|c|}
      \hline \hline
      Target & Reaction & Thickness & Density & Purity & Radiation length & Length in $X_{0}$\\
             &          & (mm)  & (g/cm$^{3}$) & ($\%$)&  $X_{0}$ (g/cm$^{2}$)&  \\ \hline \hline
      CH$_{2}$ & $p${\eek}$\Lambda$,$\Sigma^{0}$&5.0 & 0.90 & - & 44.8 & 1.0$\times$10$^{-2}$\\
             & $^{12}$C{\eek}$^{12}_{\Lambda}$B& & & & &\\ \hline
      H$_{2}$O & $p${\eek}$\Lambda$,$\Sigma^{0}$&5.0 & 1.00 & - & 36.1 &1.4$\times$10$^{-2}$\\
             & $^{16}$O{\eek}$^{16}_{\Lambda}$N& & & & &\\ \hline             
      $^{7}$Li & $^{7}$Li{\eek}$^{7}_{\Lambda}$He &3.9 & 0.54 & 99.9 & 82.8 &2.5$\times$10$^{-3}$\\ \hline
      $^{9}$Be & $^{9}$Be{\eek}$^{9}_{\Lambda}$Li &1.0 & 1.85 & 100.0 & 65.2 & 2.9$\times$10$^{-3}$\\ \hline
      $^{10}$B & $^{10}$B{\eek}$^{10}_{\Lambda}$Be &0.3 & 2.16 & 99.9 &  49.2& 1.1$\times$10$^{-3}$ \\ \hline
      $^{12}$C & $^{12}$C{\eek}$^{12}_{\Lambda}$B & 0.5 & 1.75 & 98.89 & 42.7 &2.0$\times$10$^{-3}$\\ 
               & & & & ($^{13}$C:1.11)& & \\ \hline
      $^{52}$Cr & $^{52}$Cr{\eek}$^{52}_{\Lambda}$V & 0.2& 7.15 & 99.9 & 15.3 &8.8$\times$10$^{-3}$\\ 
      \hline \hline
    \end{tabular}
  \end{center}
\end{table*}

\begin{figure}[!htbp]
  \begin{center}
    \includegraphics[width=8.6cm]{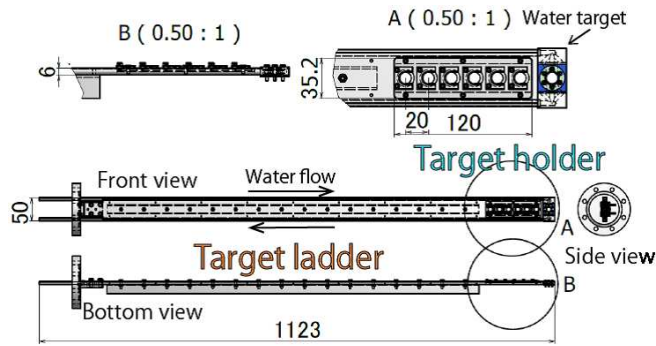}
    \caption{A schematic of the target ladder.
      The target holder which has some frames to hold 
      targets was put on the ladder.
      The target holders with different target materials
      were exchanged two times during the experiment.
      The dimensions are in mm.}
    \label{fig:target_ladder}
  \end{center}
\end{figure}

\begin{figure}[!htbp]
  \begin{center}
    \includegraphics[width=8.6cm]{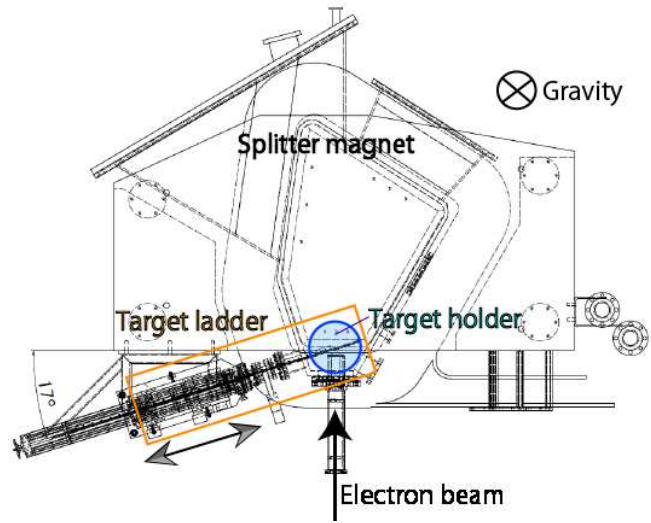}
    \caption{A schematic of the SPL and the target system.
      The target ladder was inserted at the SPL entrance with 
      the angle of seventeen degrees between the beam direction and
      the normal of the target surface.}
    \label{fig:target_spl}
  \end{center}
\end{figure}

A target holder which had several frames to fix the solid targets 
was attached to a target ladder as shown in Fig.~\ref{fig:target_ladder}.
The target ladder, made primarily of aluminum,
was inserted at the SPL entrance with the normal to the target surface
at an angle of seventeen degrees with respect to
the beam direction, as shown in Fig.~\ref{fig:target_spl}.
The position of the target holder was controlled 
by remotely sliding the target ladder in order to change 
the target intercepting the beam.
The target ladder required cooling as the targets were heated
by the intense electron beam.
For example, 
the heat deposit was estimated to be approximately 8~W
for 50-$\mu $A beam on a 0.1-g/cm$^{2}$ carbon target.
Thus, water at the room temperature ($\sim$ 25 $^{\circ}$C)
continuously flowed along the edge of target ladder
to remove the heat from the solid targets. 
Moreover, the water cell which consisted of 25-$\mu$m 
Havar foil in back and front was fabricated at the end of target ladder
in order to use water as a target.
Havar is a non-magnetic cobalt-base alloy which exhibits high strength~\cite{cite:havar}.


Prior to the experiment, 
the maximum beam current for each target was estimated
taking into account the melting point and heat conduction~\cite{cite:shichijo}
by using ANSYS~\cite{cite:ansys},
a three-dimensional finite element method software package.
As a result, the expected maximum beam currents on the 0.1-g/cm$^{2}$ thick targets
were obtained as shown in Table~\ref{tab:beammax}. 
The beam intensities in the experiment
were determined according to the above simulation results.
\begin{table*}[!htbp]
  \begin{center}
    \caption{The expected maximum beam current for each target with 
      a thickness of 0.1~g/cm$^{2}$~\cite{cite:shichijo}. 
    }
    \label{tab:beammax}
    \begin{tabular}{|c|c|c|c|}
      \hline \hline
      Target & Melting point~\cite{cite:rika} & Expected maximum & Beam current \\ 
       & (K) &  temperature (K) & ($\mu$A) \\ \hline 
      $^{7}$Li & 454  & 386 (Result without rastering) & 30 \\
      $^{10}$B & 2349 & 970 & 50 \\
      $^{12}$C & 4098  (Sublimation point) & 521 & 50 \\ 
      $^{52}$Cr& 2180 & 988 & 50 \\ \hline \hline
    \end{tabular}
  \end{center}
\end{table*}

\subsection{TUL}
\label{sec:tul}
For the trigger logic in the experiments,
the Tohoku universal logic module 
(TUL, TUL-8040)~\cite{cite:yokota},
a programmable logic module, was developed
to reduce the number of NIM modules and cables needed. 
A field programmable gate array (FPGA)
of ALTERA~\cite{cite:altera} APEX 20K series was
mounted on TUL. 
The major specifications of this module are summarized in Table~\ref{tab:tul}.
The introduction of the TUL made it possible to have an
on-line grouping trigger as described in Sec.~\ref{sec:trig},
and it reduced the risks of missed connections among
hardware circuits.
\begin{table}[!htbp]
  \begin{center}
    \caption{Major specifications of TUL.}
    \label{tab:tul}
    \begin{tabular}{c|c}
      \hline \hline
      \multicolumn{2}{c}{FPGA} \\ \hline
      Product & ALTERA APEX 20K (EP20K300E) \\
      Maximum gates & 728,000 \\
      Logic elements & 11,520 \\ \hline \hline
      \multicolumn{2}{c}{I/O} \\ \hline \hline
      Input & NIM: 16~ch \\
            & ECL: 64~ch \\ \hline
      Output & NIM: 8~ch \\
             & ECL: 32~ch \\ \hline
      Internal clock & 33~MHz \\ 
      \hline \hline
    \end{tabular}
  \end{center}
\end{table}

\subsection{Data-taking trigger}
\label{sec:trig}
A logical condition of the data-taking trigger 
for physics run (COIN$_{{\rm trigger}}$) consisted of: 
\begin{eqnarray}
  \label{eq:cointrigger}
  {\rm COIN}_{{\rm trigger}} = {\rm HKS}_{{\rm trigger}} \otimes {\rm HES}_{{\rm trigger}} 
\end{eqnarray}
where HKS$_{{\rm trigger}}$ and HES$_{{\rm trigger}}$
are trigger conditions in HKS and HES, respectively, to be explained below.

\begin{figure}[!htbp]
  \begin{center}
  \includegraphics[width=8.6cm]{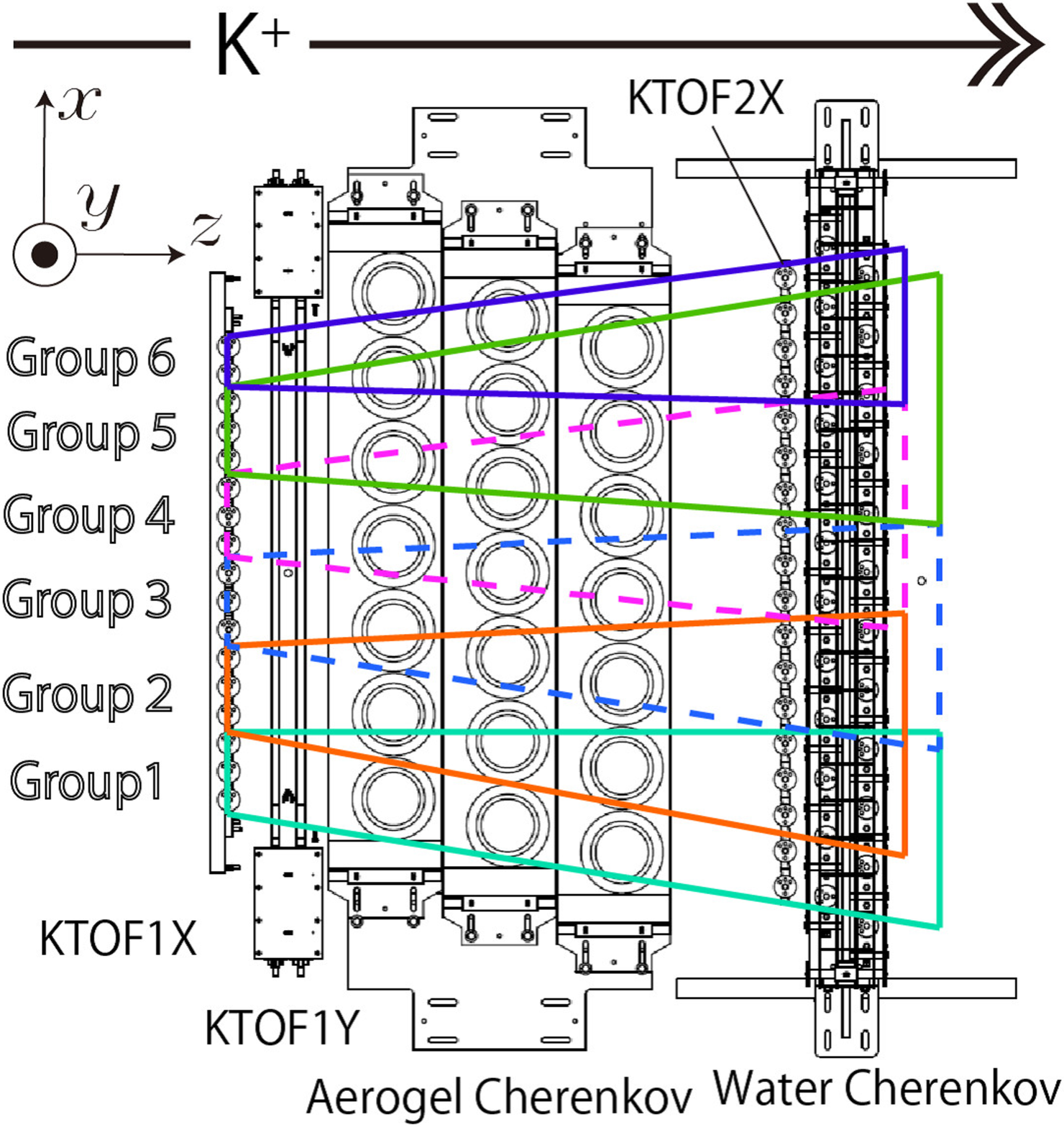}
  \caption{A schematic of the HKS-detector grouping used for the HKS kaon trigger.
    The HKS detectors were divided into six groups
    taking into account the optics of HKS. }
  \label{fig:grouping}
  \end{center}
\end{figure}
\begin{figure}[!htbp]
  \begin{center}
  \includegraphics[width=8.6cm]{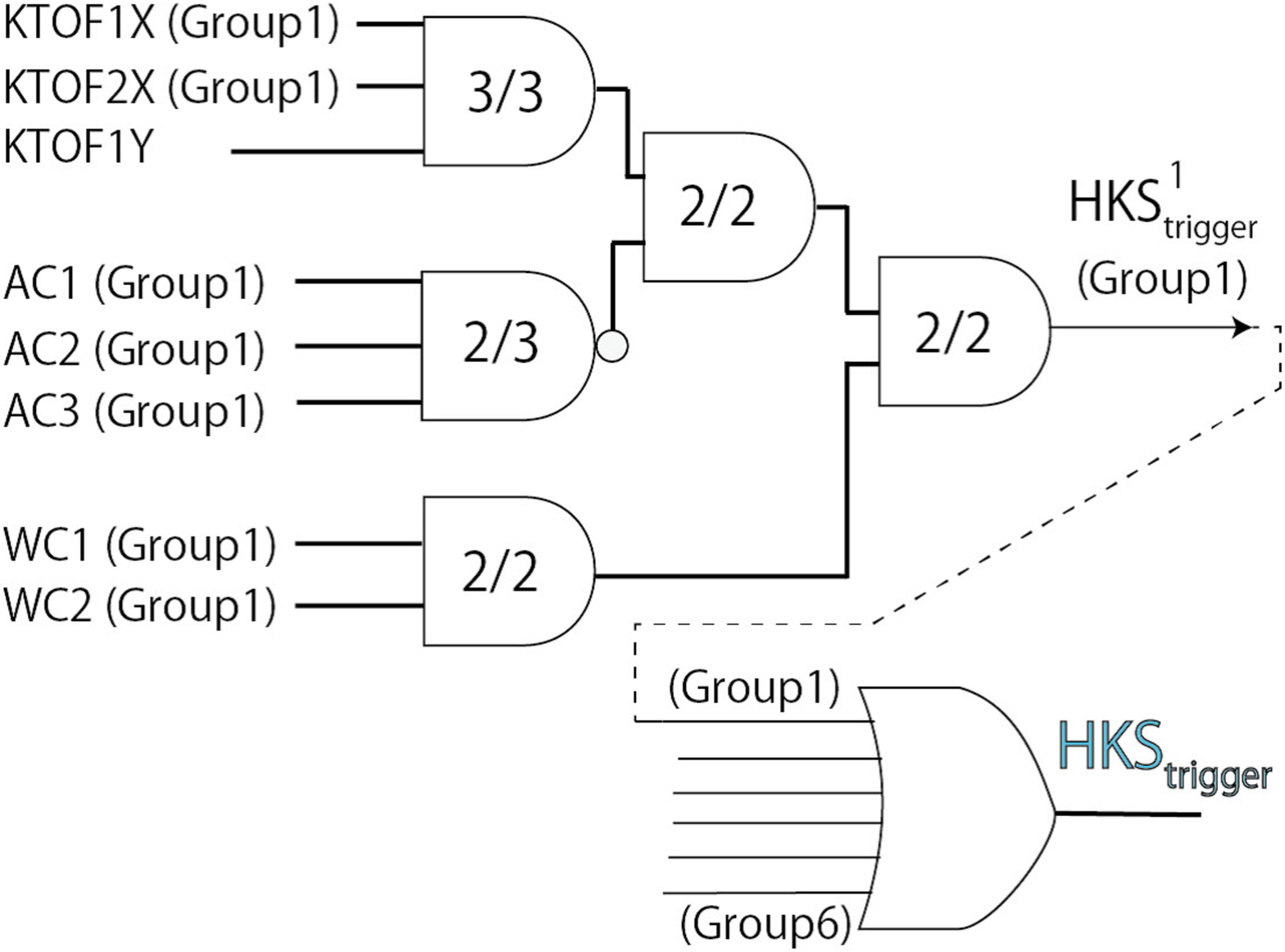}
  \caption{Diagram of logic circuit of the HKS trigger.
    The HKS detectors were divided into six groups as shown 
    in Fig.~\ref{fig:grouping}, taking into account the HKS optics.
    The trigger was made in each group (HKS$^{i}_{{\rm trigger}}$), and 
    added logically (OR) being HKS$_{{\rm trigger}}$.}
  \label{fig:hkstrigger}
  \end{center}
\end{figure}

For the HKS trigger, the detectors were divided 
into six groups taking into account
the HKS optics as shown in Fig.~\ref{fig:grouping}.
A detector combination for each group was
determined by the Monte Carlo simulation in
order to minimize the $K^{+}$-overkill ratio
as well as background contamination.
A trigger was made for each group (HKS$^{i}_{{\rm trigger}}$) 
and logically added (OR) 
to form the HKS trigger (HKS$_{{\rm trigger}}$): 
\begin{eqnarray}
  {\rm HKS}_{{\rm trigger}} = \sum^{6}_{i=1} {\rm HKS}^{i}_{{\rm trigger}}, 
\end{eqnarray}
where $i$ is the group number (grouping trigger).
Particles which were not in the HKS optics 
could be reduced by the grouping trigger.
The HKS trigger of the $i^{{\rm th}}$ group consisted of the following logical condition:
\begin{eqnarray}
  \label{eq:hks_trigger}
  {\rm HKS}^{i}_{ {\rm trigger} } =  
  {\rm CP}^{i}_{{\rm trigger}} \otimes \overline{ {\rm AC}^{i} }  \otimes {\rm WC}^{i}
\end{eqnarray}
where,    
\begin{eqnarray}
  \label{eq:cp_trigger}
  {\rm CP}^{i}_{{\rm trigger}} = {\rm KTOF1X}^{i} \otimes {\rm KTOF1Y} \otimes {\rm KTOF2X}^{i}.
\end{eqnarray}
The CP$_{{\rm trigger}}$ shown in Sec.~\ref{sec:pardet} is
defined by $\sum^{6}_{i=1} {\rm CP}^{i}_{{\rm trigger}}$.
The AC$^{i}$ and WC$^{i}$ in Eq.~(\ref{eq:hks_trigger}) denote 
$[$(AC1 $\otimes$ AC2)$^{i}$ $\oplus$ (AC2 $\otimes$ AC3)$^{i}$ $\oplus$ (AC3 $\otimes$ AC1)$^{i}]$,  
and (WC1 $\otimes$ WC2)$^{i}$, respectively. 
The overline on AC$^{i}$ indicates that 
the AC$^{i}$ was used as a veto for $\pi ^{+}$ suppression.
The logic circuit of the HKS trigger is shown in Fig.~\ref{fig:hkstrigger}.
This complicated trigger condition was
realized by the introduction of the TUL (Sec.~\ref{sec:tul}). 

The HES electron trigger was simpler than the HKS kaon trigger.
The logic condition of the HES$_{{\rm trigger}}$ was as follows:
\begin{eqnarray}
  \label{eq:hes_trigger}
  {\rm HES}_{ {\rm trigger} } =  {\rm ETOF1} \otimes {\rm ETOF2}, 
\end{eqnarray}
where 
\begin{eqnarray}
  {\rm ETOF1} &=& \sum^{29}_{j=1}{\rm ETOF1}^{j}, \\
  {\rm ETOF2} &=& \sum^{29}_{j=1}{\rm ETOF2}^{j}. \\
  &&(j:{\rm segment \hspace{0.1cm} number})\nonumber 
\end{eqnarray}

The typical counting rate for each data set
is summarized in Table~\ref{tab:trigger_rate}.
It is noted that the HES-collimator setting for 
the $^{52}$Cr-target data was different from that
of the other targets, and thus, HES$_{{\rm trigger}}$ and COIN$_{{\rm trigger}}$
for the $^{52}$Cr target cannot be directly compared to the other targets.
\begin{table*}[!htbp]
  \begin{center}
    \caption{Typical trigger rate for each target 
      in the JLab E05-115 experiment.}
    \label{tab:trigger_rate}
    \begin{tabular}{|cc|cccc|}
      \hline \hline
      Target & Beam current & \multicolumn{4}{c|}{Rate~[kHz]} \\ 
      (g/cm$^{2}$)& ($\mu$A) & CP$_{{\rm trigger}}$ & HKS$_{{\rm trigger}}$ 
             & HES$_{{\rm trigger}}$ & COIN$_{{\rm trigger}}$  \\ \hline
      CH$_{2}$ (0.451) & 2.0 & 220 & 1.8 & 1200 & 0.10  \\
      H$_{2}$O (0.500) & 2.8 & 1100& 20 & 1500 & 1.50  \\
      $^{7}$Li (0.208) & 35  & 540 & 7.3 & 2200 & 0.96 \\
      $^{9}$Be (0.188) & 40 & 710 & 1.0 & 2500 & 1.50  \\
      $^{10}$B (0.056) & 40 & 190 & 2.0 & 1600 & 0.17 \\
     $^{12}$C (0.088)  & 35 & 630 & 7.9 & 2300 & 1.30 \\
      $^{52}$Cr (0.134) & 8.0 & 980 & 11 & 2500 & 1.80  \\ \hline \hline
    \end{tabular}
  \end{center}
\end{table*}

\section{Analysis}
\label{sec:analysis}
In this section, some of the important analysis steps are
described: missing mass reconstruction, $K^{+}$ identification,
event selection for $e^{\prime}K^{+}$ coincidence,
and energy scale calibration.
Background particles which were not in the optics were detected
in addition to expected backgrounds of protons and $\pi^{+}$s in the HKS.
The origin of the backgrounds and an event selection method which we
applied to eliminate them in off-line analysis are discussed in
Sec.~\ref{sec:eeplus}.

\subsection{Missing mass reconstruction}
\label{sec:mm_recon}
The position and angle of a $K^{+}$ and scattered electron at 
the reference planes
in the magnetic spectrometers were measured by the particle detectors.
This information was converted to momentum vectors at 
the target position with backward transfer matrices (BTM)
of the optical systems for the SPL $+$ HES and SPL $+$ HKS, respectively,
in order to reconstruct a missing mass $M_{{\rm HYP}}$.
The $M_{{\rm HYP}}$ is be calculated as follows: 
\begin{eqnarray}
  M_{{\rm HYP}} &=& \Bigr[ E_{{\rm HYP}}^{2} - \vec{p}_{{\rm HYP}}^{\hspace{0.1cm}2} \Bigl]^{\frac{1}{2}} \nonumber \\
  &=& \Bigr[ ( E_{e}+M_{{\rm target}}-E_{K}-E_{e^{\prime}} )^{2} \nonumber \\
  && - ( \vec{p}_{e} - \vec{p}_{K} - \vec{p}_{e^{\prime}} )^{2} \Bigl]^{\frac{1}{2}} \nonumber \\
  &=& \Bigl[ ( E_{e}+M_{{\rm target}}-E_{K}-E_{e^{\prime}} )^{2} \nonumber \\
  && - (p_{e}^{2}+p_{K}^{2}+p_{e^{\prime}}^{2} \nonumber \\
  && - 2p_{e}p_{K}\cos{\theta_{eK}} 
  - 2p_{e}p_{e^{\prime}}\cos{\theta_{ee^{\prime}}} \nonumber \\
  && + 2p_{e^{\prime}}p_{K} \cos{ \theta_{ e^{\prime}K } } ) 
  \Bigr]^{\frac{1}{2}} \label{eq:mm0}
\end{eqnarray}
where $E$, $\vec{p}$ and $M_{{\rm target}}$ are the energy, momentum vectors, 
and mass of the target nucleus. 
The beam-momentum vector $\vec{p}_{e}$ was
precisely determined by the accelerator 
($\Delta E_{e} / E_{e} \leq 10^{-4}$ (FWHM), emittance of 2~$\mu $m$\cdot$mrad). 
Therefore, only the momentum vectors of $K^{+}$
and scattered electron ($\vec{p}_{K}$ and $\vec{p}_{e^{\prime}}$)
are necessary to deduce the missing mass in the experiment.
Once $M_{{\rm HYP}}$ is obtained, 
the $\Lambda$ binding energy $B_{\Lambda}$ can be calculated by: 
\begin{eqnarray}
  \label{eq:binding_energy}
  B_{\Lambda} (^{{\rm A}}_{\Lambda}Z) = M (^{{\rm A}-1}Z) 
  + M_{\Lambda} - M_{{\rm HYP}} (^{{\rm A}}_{\Lambda}Z)
\end{eqnarray}
where $Z$ denotes the proton number, and 
$M (^{{\rm A}-1}Z)$ and $M_{\Lambda}$ 
are the rest masses of a core nucleus at the ground state and a $\Lambda$.

The BTM ($M^{{\rm R2T}}$),
which converts the position and angle of a particle at the reference plane 
to the momentum vector at the target, 
for each optical system, SPL $+$ HES and SPL $+$ HKS, is written as:
\begin{eqnarray}
  \label{eq:mat_1}
  \left(
    \begin{array}{c}
      x_{{\rm T}} \\
      x^{\prime}_{{\rm T}} \\
      y_{{\rm T}} \\
      y^{\prime}_{{\rm T}} \\
      p \\
    \end{array}
  \right)
  &=& M^{{\rm R2T}}
  \left(
    \begin{array}{c}
      x_{{\rm RP}} \\
      x^{\prime}_{{\rm RP}} \\
      y_{{\rm RP}} \\
      y^{\prime}_{{\rm RP}} \\
      x_{{\rm RP}}^{2} \\
      x_{{\rm RP}}x^{\prime}_{{\rm RP}}\\
      \vdots
    \end{array}
  \right)
\end{eqnarray}
where $x$, $y$, $x^{\prime}$ ($\equiv \frac{p_{x}}{p_{z}}$), 
$y^{\prime}$ ($\equiv \frac{p_{y}}{p_{z}}$) 
are the positions and angles at the reference plane (subscript of RP)
and the target point (subscript of T), 
and $p$ is the momentum. 
For an initial BTM calculation, $x_{{\rm T}}$, $y_{{\rm T}}$ were assumed to be zero 
as the spatial size of electron beam on the target point
was negligibly small (typically $\sigma \simeq 100$~$\mu $m),
although the beam was rastered only for low melting-point targets,
polyethylene and $^{7}$Li,
as shown in Sec.~\ref{sec:btm_displacement}.
The variables, $x^{\prime}_{{\rm T}}$, $y^{\prime}_{{\rm T}}$ and $p$ 
in Eq.~(\ref{eq:mat_1}) are written as $n^{{\rm th}}$ order polynomial functions 
as follows:
\begin{eqnarray}
  x^{\prime}_{{\rm T}} &=& 
  \sum^{n}_{a+b+c+d=0} C_{x}(a,b,c,d) 
  (x_{{\rm RP}})^{a} (x^{\prime}_{{\rm RP}})^{b} \nonumber \\
  && \times (y_{{\rm RP}})^{c} (y^{\prime}_{{\rm RP}})^{d}, \label{eq:r2t_1} \\
  \label{eq:r2t_2}
  y^{\prime}_{{\rm T}} &=& 
  \sum^{n}_{a+b+c+d=0} C_{y}(a,b,c,d) 
  (x_{{\rm RP}})^{a} (x^{\prime}_{{\rm RP}})^{b} \nonumber \\
  && \times (y_{{\rm RP}})^{c} (y^{\prime}_{{\rm RP}})^{d}, \\
  \label{eq:r2t_3}
  p &=& 
  \sum^{n}_{a+b+c+d=0} C_{p}(a,b,c,d) 
  (x_{{\rm RP}})^{a} (x^{\prime}_{{\rm RP}})^{b} \nonumber \\
  && \times (y_{{\rm RP}})^{c} (y^{\prime}_{{\rm RP}})^{d}
\end{eqnarray}
where $C_{x,y,p}(a,b,c,d)$ are elements of $M^{{\rm R2T}}$.

The initial BTMs were obtained in the full-modeled Monte 
Carlo simulations. 
Magnetic field maps, that were used in the Monte Carlo simulation, 
were calculated by Opera3D (TOSCA).
Models of SPL $+$ HKS and SPL $+$ HES were separately prepared, 
taking into account realistic geometrical information.
Then, particles were randomly generated at the target point
with uniform distributions over momentum and angular ranges
in order to obtain corresponding position and angular information
at the reference planes.
The initial BTMs were obtained with a fitting algorithm
of the singular-value decomposition by inputting the above information
at the reference planes and target.
The obtained BTMs were not perfect for describing
the real optics of our spectrometer systems 
due to 
imperfections of the simulation models. 
In fact, momentum resolutions of 
$\Delta p/p = 10^{-3}$--$10^{-2}$ (FWHM) could be 
achieved with the initial BTMs, 
although our goal was $\Delta p/p \simeq 2 \times 10^{-4}$ (FWHM).
In addition, an energy scale had not been calibrated at this
initial stage.
Therefore, the initial BTMs needed to be optimized.
We optimized the BTMs by a sieve-slit analysis~\cite{cite:12LB}
and a semi-automated optimization program
as described in Sec.~\ref{sec:energy_calib}.

The required computational cost of
the optimization process increases as the polynomial order
$n$ is increased.
For $n=3$, 210 elements
[35 (matrix elements) $\times 3$ ($x^{\prime}$, $y^{\prime}$, $p$) elements for HES, 
and similarly $35 \times 3$ elements for HKS] have to be optimized.
In contrast, for $n=6$, 1260 elements
need to be optimized and their optimizations 
require much more computation.
In the optical simulation, 
it was found that complexities of $n\geq6$ are 
needed for both SPL $+$ HKS and SPL $+$ HES systems 
to achieve our goal, $\Delta p/p \simeq 2\times 10^{-4}$ FWHM.
In the present analysis, therefore,
the complexities of $n=6$ were chosen
taking also into account 
the computational cost in the optimization process. 

\subsection{$K^{+}$ identification}
\label{kidana}
A $K^{+}$ identification (KID) was essential in both 
on-line (data-taking trigger; Sec.~\ref{sec:trig}) 
and off-line (analysis).
Major background particles in HKS were $\pi^{+}$s and protons.

On-line KID was performed with 
a combination of two types of {\cerenkov} detectors using
radiation media of pure water (refractive index of 1.33) 
and aerogel (refractive index of 1.05), as shown in Sec.~\ref{sec:trig}.
To avoid over-cutting of $K^{+}$ at the trigger level,
the trigger thresholds for water and aerogel {\cerenkov} detectors were 
set to be loose, maintaining a DAQ efficiency that was high enough ($>90\%$). 
Thus, some $\pi^{+}$s and protons remained in data, but
were rejected in the off-line analysis.

\begin{figure}[!htbp]
  \begin{center}
    \includegraphics[width=8.6cm]{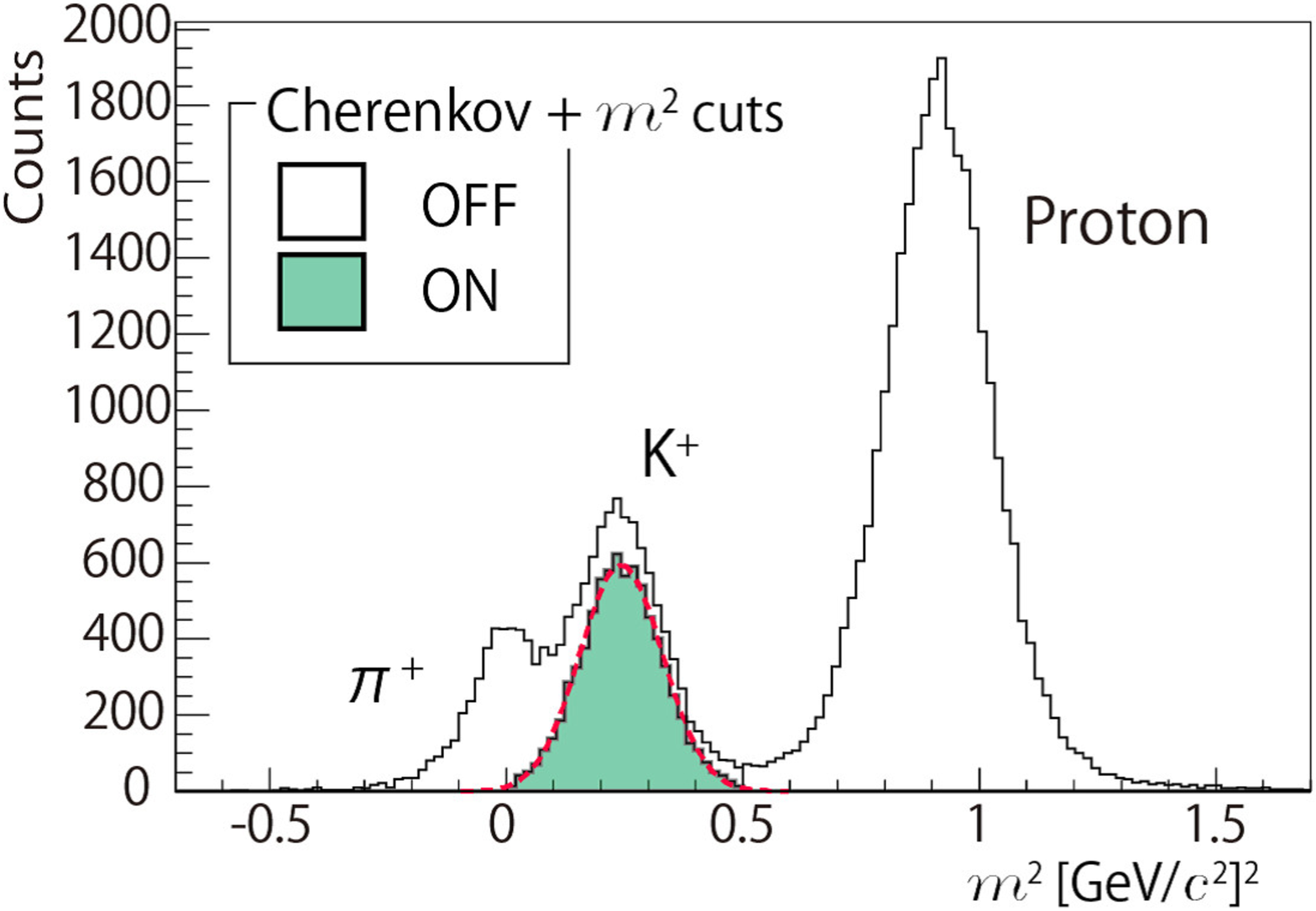}
    \caption{A typical mass squared ($m^{2}$) distribution of data 
      with the 0.451~g/cm$^{2}$ polyethylene target.
      Colored spectrum shows the $m^{2}$ distribution 
      when event selections of the number of photoelectrons 
      in the aerogel and water {\cerenkov} detectors were applied
      as shown in Fig.~\ref{fig:cherenkov_distribution}.
      $K^{+}$s were clearly separated by the
      event selections by {\cerenkov} and $m^{2}$ information.}
    \label{fig:mass_square}
  \end{center}
\end{figure}
The off-line KID was done 
by using the number of
photoelectrons (NPE) in the {\cerenkov} detectors, 
and reconstructed mass squared of the particles. 
The mass squared ($m^{2}$) was calculated by:
\begin{eqnarray}
  \label{eq:msq}
  m^{2} = p^{2} \Bigl( \frac{1}{\beta^{2}} - 1 \Bigr)
\end{eqnarray}
where $\beta$ 
is the velocity factor obtained 
by TOF and path-length measurements,
and $p$ is the particle momentum reconstructed 
by the BTM as shown in Sec.~\ref{sec:mm_recon}.
Figure~\ref{fig:mass_square} shows a typical mass squared distribution 
of data with the 0.451~g/cm$^{2}$ polyethylene target.

\begin{figure}[!htbp]
  \begin{center}
    \includegraphics[width=8.6cm]{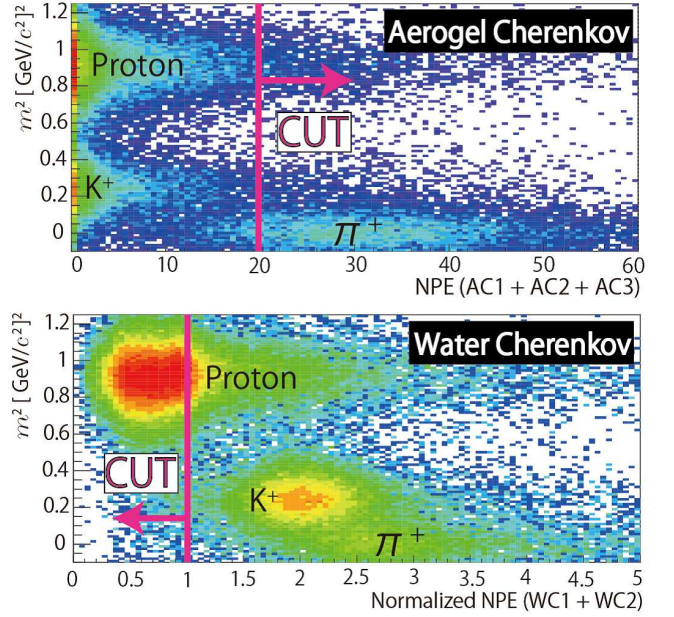}
    \caption{Correlations between the number of 
      photoelectrons (NPE) and $m^{2}$ in AC (sum of three layers) 
      and WC (sum of two layers).
      $\pi^{+}$s could be separated from 
      $K^{+}$s and protons by a cut of NPE detected by AC,
      as represented by a solid line in the top panel.
      Similarly,
      protons could be distinguished from 
      $\pi^{+}$s and $K^{+}$s by a cut on the normalized NPE
      detected by the WC, as represented by a solid line in the bottom panel.}
    \label{fig:cherenkov_distribution}
  \end{center}
\end{figure}

The top panel of Fig.~\ref{fig:cherenkov_distribution} shows 
a typical correlation between $m^{2}$ and NPE (sum of three layers)
in the aerogel {\cerenkov} detector.
The most probable value of summed NPE for $\pi^{+}$ was at about 30, 
and those of $K^{+}$ and proton were at about zero. 
Thus, the $\pi^{+}$s could be separated 
from $K^{+}$s and protons by applying a cut of NPE
as represented by a solid line in Fig.~\ref{fig:cherenkov_distribution}.
We used two types of water {\cerenkov} detectors 
(type A~\cite{cite:taniya} and B~\cite{cite:okayasu_d})
by which detection capabilities of a {\cerenkov}
radiation were different.
Main differences between these two types were
reflection materials and choice of photo multiplier tubes.
The type A was able to detect two times larger NPE
than the type B, and 
the type A was installed for higher momentum side
where better capability of proton-$K^{+}$ separation was required~\cite{cite:gogami}.
In the analyses, the most probable value of NPE for $K^{+}$ in 
a layer of the water {\cerenkov} detector was normalized to unity.
A bottom panel of Fig.~\ref{fig:cherenkov_distribution} shows 
a typical correlation between $m^{2}$ and normalized NPE (sum of two layers)
in the water {\cerenkov} detector.
As with the aerogel {\cerenkov} detector, 
protons could be separated from $\pi^{+}$s and $K^{+}$s 
by a cut on the normalized NPE in the water {\cerenkov} detector, 
as represented by a solid line in Fig.~\ref{fig:cherenkov_distribution}.
The colored spectrum in Fig.~\ref{fig:mass_square} shows a
typical $m^{2}$ distribution with 
the above cuts of $\pi^{+}$s and protons by the {\cerenkov} detectors
(Fig.~\ref{fig:cherenkov_distribution}), 
and a $m^{2}$ selection of $|m^{2} - m^{2}_{K}| \leq 0.3$ where $m_{K}$ is the known mass of $K^{+}$~\cite{cite:pdg}. 
The $K^{+}$ peak in the $m^{2}$ distribution 
was fitted with a Gaussian function,  
and the width was found to be $\sigma \simeq (0.29$~GeV/$c^{2})^{2}$. 
When the off-line KID cuts were 
selected to maintain a $90\%$ $K^{+}$ survival ratio,
the total (on-line and off-line) rejection capabilities of $\pi^{+}$s and protons were
$4.7 \times 10^{-4}$ and $1.9 \times 10^{-4}$, respectively, 
for the case of a 2-$\mu $A beam on the 0.451~g/cm$^{2}$ polyethylene target~\cite{cite:gogami}.

\subsection{Event Selection for Real $e^{\prime}K^{+}$ Coincidence}
\label{sec:coincidence}
In order to find proper coincidences between 
$e^{\prime}$s and $K^{+}$s in the data, 
we defined a coincidence time ${\rm T}_{\rm coin}$ as follows:
\begin{eqnarray}
  {\rm T}_{{\rm coin}} = {\rm T}_{{\rm HKS}} - {\rm T}_{{\rm HES}}
  \label{eq:cointime}
\end{eqnarray}
where T$_{{\rm HKS}}$ and T$_{{\rm HES}}$ are reconstructed times at the target position 
in the HKS and HES, respectively.
T$_{{\rm HKS}}$ and T$_{{\rm HES}}$ were calculated by 
using the times at the TOF detectors (KTOF, ETOF), 
the path lengths between the TOF detectors and the target, 
and the velocity factors ($\beta$) of particles. 
The path lengths were derived by backward transfer matrices, 
and the velocity factors were measured by the TOF detectors. 
\label{coinana}
\begin{figure}[!htbp]
  \begin{center}
    \includegraphics[width=8.6cm]{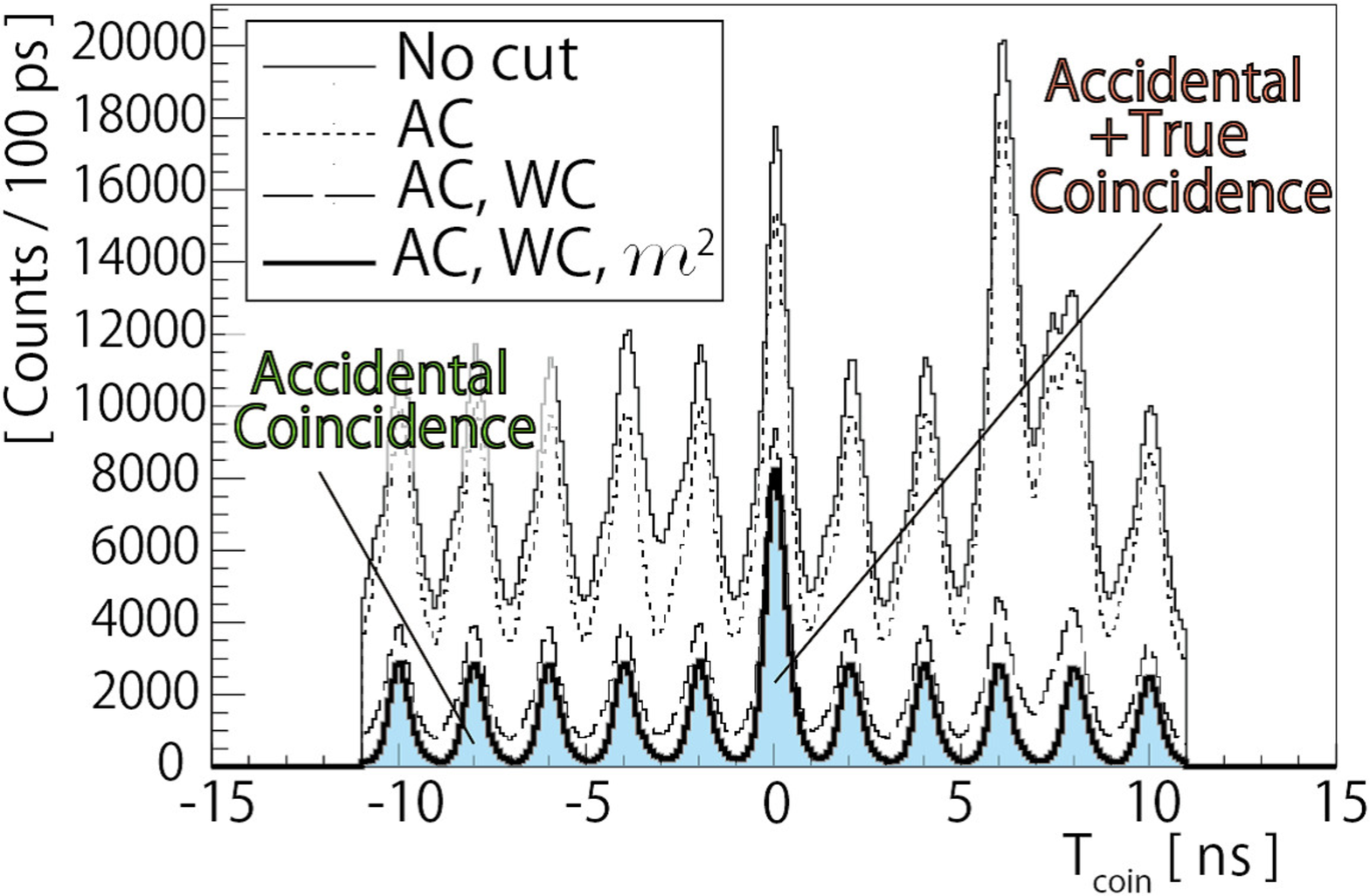}
    \caption{Coincidence time [${\rm T}_{{\rm coin}}$ in Eq.~(\ref{eq:cointime})] 
      distribution with and without off-line KID as shown in Sec.~\ref{kidana}, 
      for the case of a 2-$\mu $A beam on the 0.451-g/cm$^{2}$ polyethylene target.
      AC, WC and $m^{2}$ denote event selections for $K^{+}$ by 
      the number of photoelectrons in the aerogel and water {\cerenkov} detectors 
      (Fig.~\ref{fig:cherenkov_distribution}),
      and the reconstructed mass squared, $m^{2}$ (Fig.~\ref{fig:mass_square}). }
    \label{fig:ctime1}
  \end{center}
\end{figure}
Figure~\ref{fig:ctime1} shows a typical ${\rm T}_{{\rm coin}}$ distribution 
with and without the off-line KID as shown in Sec.~\ref{kidana}. 
The beam-bunch interval of CEBAF was 2~ns, and the beam-bunch
structure was clearly observed with a resolution of 
$\sigma \simeq 270$~ps. 
The beam bunch at ${\rm T}_{{\rm coin}}=0$~ns was enhanced 
after the off-line KID by event selections of 
the number of photoelectrons in the {\cerenkov} detectors (AC, WC) 
and the reconstructed mass squared ($m^{2}$) of particles. 
Hence, the peak at ${\rm T}_{{\rm coin}}=0$~ns contains 
events of true coincidence between $e^{\prime}$ and $K^{+}$, 
while the other peaks contain only accidental coincidence events. 
In the analyses, events of $|{\rm T}_{{\rm coin}}| \leq 1.0$~ns 
were selected as the true $e^{\prime}K^{+}$-coincidence events.


\subsection{Energy Scale Calibration}
\label{sec:energy_calib}
\begin{figure}[!htbp]
  \begin{center}
    \includegraphics[width=8.7cm]{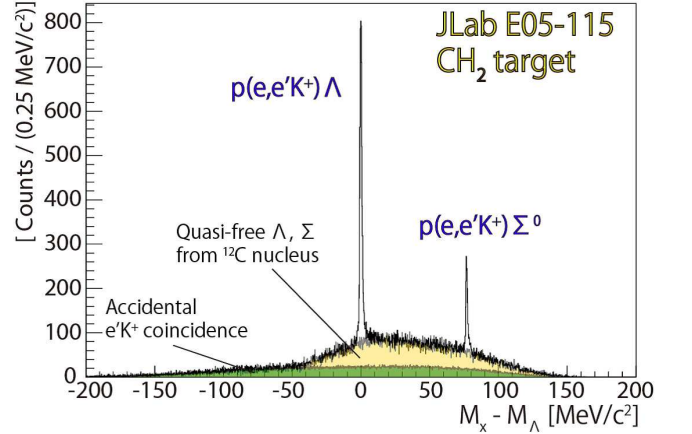}
    \caption{Missing mass spectrum for $\Lambda$ and $\Sigma^{0}$ from 
      the 0.451-g/cm$^{2}$ polyethylene target.
      There are clear peaks of $\Lambda$ and $\Sigma^{0}$ with the 
      width of about $1.5$-MeV FWHM on the top of 
      background events due to the accidental coincidence and 
      $\Lambda$/$\Sigma^{0,-}$ production from $^{12}$C nuclei in the polyethylene target. 
      These background distributions were able to be obtained from the real data.
    }
    \label{fig:lambda_mm}
  \end{center}
\end{figure}

The energy scale calibration was performed by optimizing the
BTMs of our magnetic spectrometer systems~\cite{cite:12LB}.
For the BTM optimization, we used events of $\Lambda$ and $\Sigma^{0}$ 
from the 0.451-g/cm$^{2}$ polyethylene target, 
and those of the ground state of $^{12}_{\Lambda}$B 
from the 0.088-g/cm$^{2}$ natural carbon target.
Figure~\ref{fig:lambda_mm} shows the missing mass spectrum from 
the polyethylene target showing clear peaks of $\Lambda$ and 
$\Sigma^{0}$ on the top of widely distributed background events. 
These backgrounds originate from the accidental 
coincidence and the $\Lambda$/$\Sigma^{0,-}$ production from $^{12}$C nuclei.
The distribution of the accidental background 
was obtained by selecting events in off-time gates (Sec.~\ref{sec:coincidence}).
Moreover, it can also be obtained with a negligibly small statistical uncertainty
by the mixed event analysis as applied
for analyses of $\Lambda$ hypernuclei~\cite{cite:7LHe,cite:12LB,cite:7LHe_2}.
On the other hand, the distribution of
$\Lambda$ and $\Sigma^{0,-}$ production from $^{12}$C nuclei
was obtained from the analysis of the natural carbon data.

In the BTM-optimization process, a $\chi^{2}$ to be minimized 
was defined as follows: 
\begin{eqnarray}
  \label{eq:chi2_tuning}
  \chi^{2} = \frac{1}{3} \sum^{3}_{i=1}  w_{i} \chi^{2}_{i}
\end{eqnarray}
where 
\begin{eqnarray}
  \chi^{2}_{1} &=& \frac{1}{N_{\Lambda}} 
  \sum^{N_{\Lambda}}_{j=1}\Bigl( \frac{M^{j}_{\Lambda} 
    - M_{\Lambda}}{\sigma_{\Lambda}} \Bigr)^{2},  \\
  \chi^{2}_{2} &=& \frac{1}{N_{\Sigma^{0}}} 
  \sum^{N_{\Sigma^{0}}}_{k=1}\Bigl( \frac{M^{k}_{\Sigma^{0}} 
    - M_{\Sigma^{0}}}{\sigma_{\Sigma^{0}}} \Bigr)^{2}, \\
  \chi^{2}_{3} &=& \frac{1}{N_{{\rm 12BL}}}
  \sum^{N_{{\rm 12BL}}}_{l=1}\Bigl( \frac{M^{l}_{{\rm 12BL}} 
    - M^{fit}_{{\rm 12BL}}}{\sigma_{{\rm 12BL}}} \Bigr)^{2}.
\end{eqnarray}
The $M^{j,k,l}_{\Lambda,\Sigma^{0},{\rm 12BL}}$ are the reconstructed 
missing masses of $\Lambda$, $\Sigma^{0}$ and the ground state of $^{12}_{\Lambda}$B.
The $M_{\Lambda, \Sigma^{0}}$ are the well known masses of $\Lambda$ and 
$\Sigma^{0}$~\cite{cite:pdg}.  
The $M^{fit}_{{\rm 12BL}}$ denotes a mean value obtained by a single Gaussian fitting to 
the ground state of $^{12}_{\Lambda}$B in an iteration process of the BTM optimization.
The $\sigma_{\Lambda, \Sigma^{0}, {\rm 12BL}}$ represent
expected missing-mass resolutions estimated by the Monte Carlo simulation (Sec.~\ref{sec:results}). 
The $N_{\Lambda, \Sigma^{0}, {\rm 12BL}}$ and $w_{\Lambda, \Sigma^{0}, {\rm 12BL}}$ are 
the number of events and weights for $\Lambda$, $\Sigma^{0}$ and the ground 
state of $^{12}_{\Lambda}$B.

\begin{figure*}[!htbp]
  \begin{center}
    \includegraphics[width=13cm]{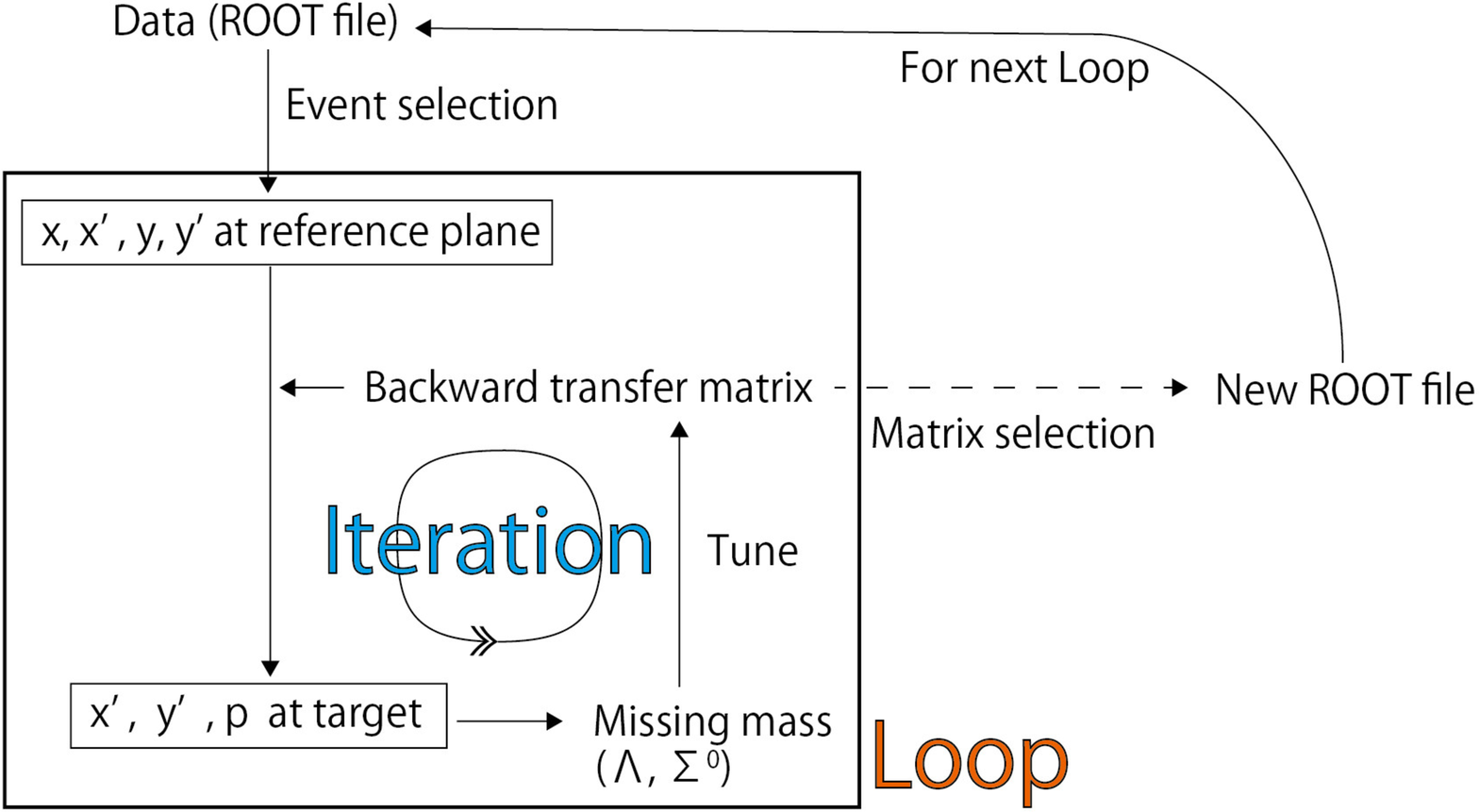}
    \caption{Flow chart of the backward transfer matrix optimization.}
    \label{fig:tuning}
  \end{center}
\end{figure*}
Figure~\ref{fig:tuning} shows the flow chart of the BTM optimization.
At first, event samples for the tuning 
were selected with a certain condition.
The BTM elements for 
angles [Eq.~(\ref{eq:r2t_1}) and Eq.~(\ref{eq:r2t_2})] and
momenta [Eq.~(\ref{eq:r2t_3})] were alternately optimized 
in the iterative optimization process.
We used an weight ratio of 
$w_{\Lambda}:w_{\Sigma^{0}}:w_{{\rm 12BL}}=1:1:0$
when the angular elements were optimized.
The ground-state events of $^{12}_{\Lambda}$B
were not used in the angular 
element optimization since kinamatically the angular contributions 
of the hypernucleus to the missing mass are
negligibly small relative to the hyperons (angular contributions for $\Lambda$ 
are approximately ten times larger than those for $^{12}_{\Lambda}$B;
Table~\ref{tab:mm_contribution}).
For the momentum element optimization, on the other hand, 
the weight ratio of $w_{\Lambda}:w_{\Sigma^{0}}:w_{{\rm 12BL}}=1:1:2$ 
was used. 
It is noted that the $M^{fit}_{{\rm 12BL}}$ was not a fixed value, but 
a mean value of fitting result by a Gaussian-function in each iteration. 
Thus, in the BTM optimization, the energy scale was calibrated 
by events of $\Lambda$ and $\Sigma^{0}$, 
and the ground-state events of $^{12}_{\Lambda}$B predominantly contributed
to improving the energy resolution. 
New BMTs used for the next process (loop; Fig.~\ref{fig:tuning}) 
were selected according to checks of missing-mass resolutions and 
peak positions of $\Lambda$, $\Sigma^{0}$ and 
the $^{12}_{\Lambda}$B-ground state after a
number of tuning iterations.
Event samples for each next loop were selected with 
missing masses reconstructed by the new BTMs.
The above BTM optimization was repeated until 
the missing mass resolutions achieved
the values expected from simulations.
The above process was essentially automated.
At some points, however, event-selection conditions
were adjusted by hand,
depending on the energy resolution,
in order to improve $S/N$ of events used for the tuning process. 

Systematic errors which originated from the above BTM-optimization process 
needed to be estimated carefully as the BTM optimization 
mainly determines the accuracy of the binding energy ($B_{\Lambda}$) 
and excitation energy ($E_{\Lambda}$) of a $\Lambda$ hypernucleus.
In order to estimate the achievable energy accuracy, 
we performed a full-modeled Monte Carlo simulation with dummy data.
The dummy data were generated, taking into account realistic $S/N$ and yields
of $\Lambda$, $\Sigma^{0}$ and hypernuclei.
Initially the BTMs were perfect in the simulation. 
Therefore, the BTMs were distorted so as to reproduce broadening and shifts in 
the missing mass spectra as much as those for the real data. 
Then, the distorted BTMs were optimized by the exactly same code as that for the real data, 
and the obtained energies ($B_{\Lambda}$, $E_{\Lambda}$) were compared with assumed energies. 
The above procedure was tested several times by using different sets of dummy data and BTMs.
As a result, it was found that $B_{\Lambda}$ and $E_{\Lambda}$ could be 
obtained with the accuracy of $< 0.09$~MeV and $< 0.05$~MeV, respectively 
using the above calibration method.
An uncertainty of target thickness, 
which was estimated to be $5\%$ according to 
accuracy of its fabrication and thickness measurement,
is considered to be another 
major contribution to $\Delta B_{\Lambda}$.
It is noted that the target thickness uncertainty
is canceled out for the $E_{\Lambda}$ calculation. 
The energy losses of particles in each target
were evaluated by the Monte Carlo simulation, 
and used as a correction to the missing mass calculation
as shown in Sec.~\ref{sec:massoffset}.
The energy loss correction has an uncertainty
due to the target thickness uncertainty, and
it was evaluated by the Monte Carlo simulation to
be taken into account for $\Delta B_{\Lambda}$. 
Consequently, 
total systematic errors on $B_{\Lambda}$ and $E_{\Lambda}$ were estimated to be
$\pm 0.11$~MeV and $\pm0.05$~MeV, respectively.
\subsection{$e^{-}e^{+}$ background in HKS}
\label{sec:eeplus}
\begin{figure}[!htbp]
  \begin{center}
    \includegraphics[width=8cm]{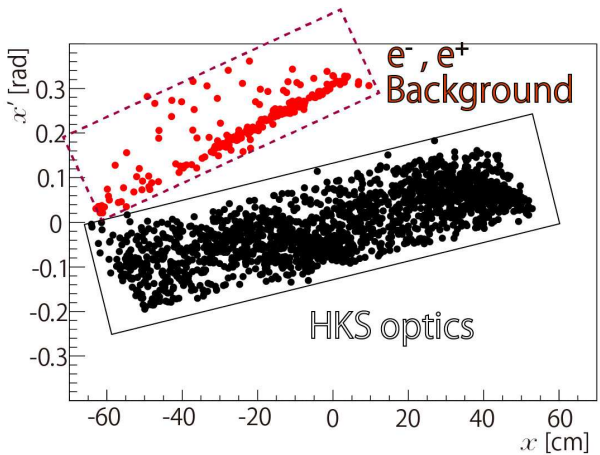}
    \caption{Distribution of $x$ versus $x^{\prime}$ at the HKS reference plane
      in the case of the $^{7}$Li target.
      Events in a solid box were in the HKS optics. 
      However, there were events which were not
      on the optics as shown in a dashed-line box. }
    \label{fig:bgtrackhks2}
  \end{center}
\end{figure}

\begin{figure}[!htbp]
  \begin{center}
    \includegraphics[width=8cm]{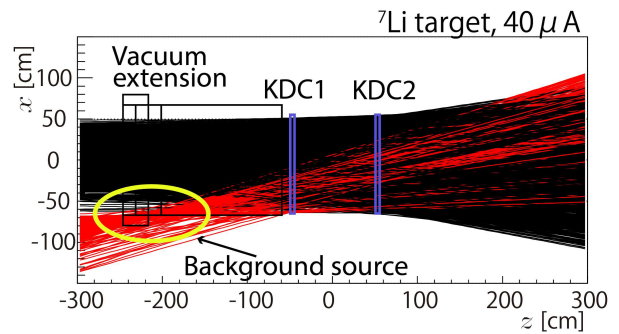}
    \caption{Reconstructed HKS tracks on the $xz$-plane 
      in the case of $^{7}$Li target for the real data.
      There were particles coming from the lower momentum side ($x<0$)
      of the vacuum-extension box
      which was made of a stainless steel SUS304.
      As a result of the Monte Carlo simulation, 
      the source of these backgrounds is considered to be 
      positrons generated in the target.
      The positrons with certain momentum and scattering angle 
      would hit the vacuum-extension box.
      It would result in a detection of positrons
      and electrons which are generated
      via the pair creation process in the walls of the vacuum-extension box.
    }
    \label{fig:bgtrackhks}
  \end{center}
\end{figure}

Electromagnetic background particles such as
electrons from the Bremsstrahlung or \moller\ scattering
were expected in the HES.  These were drastically
reduced by the tilt method as shown in 
Sec.~\ref{sec:tilt_method}. 
However, in the HKS, background events
which are attributed to $e^{-}e^{+}$ pair production
were detected in addition to the expected background hadrons ($\pi^{+}$ and proton).
Figure~\ref{fig:bgtrackhks2} shows a typical 
distribution of $x$ versus $x^{\prime}$ at the reference plane 
for the case of $^{7}$Li target.
Plots in a solid box indicate particles within the HKS optics.
Apart from the solid box, however, there are events 
constituting a band structure in the dashed-line box. 
The events were traced back toward upstream direction 
by using the particle-tracking information as 
shown in Fig.~\ref{fig:bgtrackhks}, and 
it was found that they came from secondary scattering at the low-momentum side of vacuum-extension 
box which was made of a stainless steel SUS304.
The Monte Carlo simulation reproduced such a situation.
Positrons hit the vacuum-extension box
when the positrons with the momenta of $p=0.8$--1.0~GeV/$c$ and 
the scattered angle of $\theta = 0$--2~mrad
were generated in the target via the e$^{-}$e$^{+}$-pair production process.
Then, more positrons and electrons were generated 
in the vacuum-extension box, and they were detected in HKS.
These background events were eliminated in the off-line analysis
by selecting events on the $x$ versus $x^{\prime}$ histogram as shown 
in Fig.~\ref{fig:bgtrackhks2}.


These background events were recognized during the experiment. 
However, it seemed to be hard to shield 
them physically since they passed through inside of 
the vacuum-extension box according to the on-line analysis as 
shown in Fig.~\ref{fig:bgtrackhks}.
Moreover, physical shields could be another source 
of background events. Therefore, 
we decided to take data with these background particles
in the experiment. 
The counting rate of the background for each target 
was normalized by the areal density of the target 
and beam intensity.
As a result, it was found that 
the background rate increased
with a square of the target-proton number $Z^{2}$.
It indicates that this background originated
from an electro-magnetic process. 
This is a major reason why we used
the lower beam-intensity on the $^{52}$Cr target
compared to the other lighter targets. 

The experimental setup of the 
future hypernuclear experiment E12-15-008,
which will be carried out with HKS and HRS~\cite{cite:hrs} in JLab Hall A,
is being optimized to avoid the above background events~\cite{cite:pac44}.

\section{Missig-mass resolution achieved}
\label{sec:results}
Major factors which contribute
to the missing-mass resolution are presented
in this section.
The missing-mass resolution cannot
be easily estimated by considering each contribution
separately as
they are not independent from each other.
Therefore, a full-modeled Monte Carlo simulation
was performed to investigate the achievable
resolution taking into account all of the
major factors.
A comparison between the expected mass resolutions 
and those of final results for typical hypernuclei
is shown in Sec.~\ref{sec:mm_comp_simres}.

The contributions to the missing-mass resolution
are dominated by the following sources:
1)~the intrinsic mass-resolution due to momentum and angular
resolutions of spectrometers (Sec.~\ref{sec:intmass}),
2)~mass-offset effect due to energy-loss variations in the finite volumes of the targets (Sec.~\ref{sec:massoffset}), and 
3)~production point displacements from the assumed origin
of the BTMs (Sec.~\ref{sec:btm_displacement}).
Momentum straggling in the target also contributes
to the mass-resolution, and was estimated to be
less than 50 and 150-keV FWHM for production of hypernuclei and $\Lambda$,
respectively, when they are produced at the target center.
However, it is worth noting that the 
energy-straggling contribution 
is somewhat smaller than from the above three major factors.

\subsection{Intrinsic mass resolution}
\label{sec:intmass}
The intrinsic mass resolution, which is a kinematical broadening 
due to the momentum and angular resolutions of the magnetic spectrometers 
as well as beam qualities such as a beam-energy spread, 
was estimated by the Monte Carlo simulation.
A typical value of the beam-energy spread was $\Delta E/E=3.0\times 10 ^{-5}$
which was used for the simulation.
On the other hand, the spectrometers' resolutions 
for the momentum and angle at the target point
were evaluated, as shown in Table~\ref{tab:reso_assump},
taking into account achieved resolutions of the position and angle measurements 
at the reference planes from the particle detectors.
\begin{table}[!htbp]
  \begin{center}
    \caption{Resolutions for the reconstructed momentum and angle in 
      the optical systems of SPL $+$ HKS and SPL $+$ HES obtained by the Monte Carlo simulation 
      taking into account the achieved position and angular resolutions of the particle detectors. }
  \label{tab:reso_assump}
  \begin{tabular}{|c|cc|}
    \hline
    Spectrometer  & SPL$+$HES   & SPL$+$HKS \\ 
    Particle & $e^{\prime}$ & $K^{+}$ \\ \hline
    $\Delta p$/$p$ & 4.2$\times 10^{-4}$ & 2.0$\times 10^{-4}$ \\
    $\Delta \theta$ (RMS) (mrad) & 4.0 & 0.4 \\ 
    \hline \hline
  \end{tabular}
  \end{center}
\end{table}


When the kinematical variables 
$p_{e, e^{\prime}, K}$ and $\theta_{ee^{\prime},eK,e^{\prime}K}$
are varied by $\Delta p_{e, e^{\prime}, K}$ and $\Delta \theta_{ee^{\prime},eK,e^{\prime}K}$,
the variations of $M_{{\rm HYP}}$ can be calculated as follows:
\begin{eqnarray}
  \frac{\partial M_{{\rm HYP}}}{\partial p_{e}} \Delta p_{e}
  &=& +\frac{1}{M_{{\rm HYP}}} \Bigl[ \beta_{e}(M_{{\rm target}}-E_{K}-E_{e^{\prime}}) \nonumber \\
  && + p_{e^{\prime}} \cos{\theta_{ee^{\prime}}} + p_{K} \cos{\theta_{eK}}\Bigr] \Delta p_{e}, \label{eq:mmpd1}\\
  \frac{\partial M_{{\rm HYP}}}{\partial p_{e^{\prime}}} \Delta p_{e^{\prime}}
  &=& - \frac{1}{M_{{\rm HYP}}} \Bigl[ \beta_{e^{\prime}}(M_{{\rm target}}+E_{e}-E_{K}) \nonumber \\
  && - p_{e} \cos{\theta_{ee^{\prime}}} + p_{K} \cos{\theta_{e^{\prime}K}}\Bigr] \Delta p_{e^{\prime}}, \label{eq:mmpd2}\\
  \frac{\partial M_{{\rm HYP}}}{\partial p_{K}} \Delta p_{K}
  &=& - \frac{1}{M_{{\rm HYP}}} \Bigl[ \beta_{K}(M_{{\rm target}}+E_{e}-E_{e^{\prime}}) \nonumber \\
  && - p_{e} \cos{\theta_{eK}} + p_{e^{\prime}} \cos{\theta_{e^{\prime}K}}\Bigr] \Delta p_{K},\label{eq:mmpd3}\\
  \frac{\partial M_{{\rm HYP}}}{\partial \theta_{ee^{\prime}}} \Delta \theta_{ee^{\prime}}
  &=& - \Bigl(\frac{1}{M_{{\rm HYP}}} p_{e}p_{e^{\prime}} \sin{\theta_{ee^{\prime}}}\Bigr) \Delta \theta_{ee^{\prime}},\label{eq:mmpd4}\\
  \frac{\partial M_{{\rm HYP}}}{\partial \theta_{eK}} \Delta \theta_{eK}
  &=& - \Bigl(\frac{1}{M_{{\rm HYP}}} p_{e}p_{K} \sin{\theta_{eK}}\Bigr) \Delta \theta_{eK},\label{eq:mmpd5}\\
  \frac{\partial M_{{\rm HYP}}}{\partial \theta_{e^{\prime}K}} \Delta \theta_{e^{\prime}K}
  &=&  +\Bigl(\frac{1}{M_{{\rm HYP}}} p_{e^{\prime}}p_{K} \sin{\theta_{e^{\prime}K}}\Bigr) \Delta \theta_{e^{\prime}K}.\label{eq:mmpd6}
\end{eqnarray}
The above partial differentiations were calculated event by event 
in the Monte Carlo simulation and their mean values
for the typical
reactions $p${\eek}$\Lambda$, $^{7}$Li{\eek}$^{7}_{\Lambda}$He, and $^{12}$C{\eek}$^{12}_{\Lambda}$B are 
summarized in Table~\ref{tab:mm_contribution}.
\begin{table}[!htbp]
  \begin{center}
    \caption{Mean values of the partial differentiations
      in Eqs.~(\ref{eq:mmpd1}--\ref{eq:mmpd6})
      obtained in the Monte Carlo simulation.
      Intrinsic mass resolution $\Delta M^{{\rm int}}_{{\rm HYP}}$
      which is defined by Eq.~(\ref{eq:mmint}) is shown for
      each target in the last row.}
    \label{tab:mm_contribution}
    \begin{tabular}{c|rrr}
      \hline
      & $\Lambda$ & $^{7}_{\Lambda}$He & $^{12}_{\Lambda}$B \\ \hline
      Assumed $B_{\Lambda}$ (MeV)& - & 5.5 & 11.37 \\ \hline
      $\frac{\partial M_{{\rm HYP}} }{\partial p_{e}}$ $\Bigl(\frac{{\rm keV/}c^{2}}{{\rm MeV/}c}\Bigr)$ 
      & 742 & 957 & 974 \\
      $\frac{\partial M_{{\rm HYP}} }{\partial p_{e^{\prime}}}$ $\Bigl(\frac{{\rm keV/}c^{2}}{{\rm MeV/}c}\Bigr)$ 
      & $-$747 & $-$958 & $-$975  \\
      $\frac{\partial M_{{\rm HYP}} }{\partial p_{K}}$ $\Bigl(\frac{{\rm keV/}c^{2}}{{\rm MeV/}c}\Bigr)$ 
      & $-$673 & $-$885 & $-$902 \\
      $\frac{\partial M_{{\rm HYP}} }{\partial \theta_{ee^{\prime}}}$ $\Bigl(\frac{{\rm keV/}c^{2}}{{\rm mrad}}\Bigr)$ 
      & $-$124 & $-$21 & $-$13 \\
      $\frac{\partial M_{{\rm HYP}} }{\partial \theta_{eK}}$ $\Bigl(\frac{{\rm keV/}c^{2}}{{\rm mrad}}\Bigr)$ 
      & $-$258 & $-$51 & $-$30 \\
      $\frac{\partial M_{{\rm HYP}} }{\partial \theta_{e^{\prime}K}}$ $\Bigl(\frac{{\rm keV/}c^{2}}{{\rm mrad}}\Bigr)$ 
      & 109 & 20 & 12 \\ \hline
      $\Delta M^{{\rm int}}_{{\rm HYP}}$ (keV/$c^{2}$) (FWHM) & 733 & 414 & 410 \\
      \hline
    \end{tabular}
  \end{center}
\end{table}

If all of the variables are assumed to be independent from each other, 
the intrinsic missing-mass resolution $\Delta M^{{\rm int}}_{{\rm HYP}}$ is obtained to be:
\begin{eqnarray}
  \label{eq:mmint}
  (\Delta M^{{\rm int}}_{{\rm HYP}})^{2} &=&
  \Bigl(\frac{\partial M_{{\rm HYP}} }{\partial p_{e}} \Delta p_{e} \Bigr)^{2}
  + \Bigl(\frac{\partial M_{{\rm HYP}} }{\partial p_{e^{\prime}}} \Delta p_{e^{\prime}} \Bigr)^{2} \nonumber \\
  && + \Bigl(\frac{\partial M_{{\rm HYP}} }{\partial p_{K}} \Delta p_{K} \Bigr)^{2} \nonumber  
  + \Bigl(\frac{\partial M_{{\rm HYP}} }{\partial \theta_{ee^{\prime}}} \Delta \theta_{ee^{\prime}} \Bigr)^{2} \nonumber \\
  && + \Bigl(\frac{\partial M_{{\rm HYP}} }{\partial \theta_{eK}} \Delta \theta_{eK} \Bigr)^{2} 
  + \Bigl(\frac{\partial M_{{\rm HYP}} }{\partial \theta_{e^{\prime}K}} \Delta \theta_{e^{\prime}K} \Bigr)^{2}, \nonumber \\
\end{eqnarray}
The calculated results of $\Delta M^{{\rm int}}_{{\rm HYP}}$ for the typical targets 
by using the momentum and angular resolutions shown in Table~\ref{tab:reso_assump}
are shown in the last row of Table~\ref{tab:mm_contribution}.
It should be emphasized that $\Delta M^{{\rm int}}_{{\rm HYP}}$ is
just a reference value to be compared with other effects on
the mass resolution because the variables in Eq.~(\ref{eq:mmint}) are not independent
from each other. 

\subsection{Mass offset due to the energy loss in target}
\label{sec:massoffset}
The momentum of the beam $p^{{\rm det}}_{e}$ is precisely determined by the accelerator.
However, the beam momentum at the production point $p_{e}$ is lower because of 
momentum loss in the target. 
For $e^{\prime}$ and $K^{+}$, on the other hand, 
the momenta at the production point $p_{e^{\prime},K}$ 
are higher than those measured by the spectrometers $p^{{\rm mea}}_{e^{\prime},K}$.
Therefore, 
\begin{eqnarray} 
  p^{{\rm det}}_{e} &=&  p_{e} + \delta p_{e} \\
  p^{{\rm mea}}_{e^{\prime}} &=& p_{e^{\prime}} - \delta p_{e^{\prime}} \\
  p^{{\rm mea}}_{K}  &=&  p_{K} - \delta p_{K} 
\end{eqnarray}
where $\delta p_{e,e^{\prime},K}$ are the momentum losses in the target.
The correction for the momentum loss was applied to the 
missing mass derivation as follows: 
\begin{eqnarray}
  p_{e} &=& p^{{\rm det}}_{e} - \delta p_{e}^{{\rm center}} \\
  p_{e^{\prime}} &=& p^{{\rm mea}}_{e^{\prime}}+ \delta p_{e^{\prime}}^{{\rm center}} \\
  p_{K} &=& p^{{\rm mea}}_{K} + \delta p_{K}^{{\rm center}} 
\end{eqnarray}
where $\delta p_{e,e^{\prime},K^{+}}^{{\rm center}}$ are 
the correction factors which were obtained in the 
Monte Carlo simulation assuming 
hypernuclei or hyperons produced at the target center. 
However, this correction cannot compensate for the momentum loss 
properly when the production point is displaced from the center
particularly along $z$-direction, and thus it caused missing mass broadening.
Assuming
$\delta p^{{\rm center}} = \delta p_{e}^{{\rm center}} = \delta p_{e^{\prime}}^{{\rm center}} = \delta p_{K}^{{\rm center}}$
and that the angular contribution due to the multiple scattering is small, 
the missing mass shift between production points of front and back surfaces 
of the target $\Delta M^{{\rm eloss}}_{{\rm HYP}}$ can be estimated as follows: 
\begin{eqnarray}
  \Delta M^{{\rm eloss}}_{{\rm HYP}}&\simeq&
  -\frac{\partial M_{{\rm HYP}} }{\partial p_{e}} (2\delta p_{e}^{{\rm center}}) \nonumber \\
  && -\Bigl[\frac{\partial M_{{\rm HYP}} }{\partial p_{e^{\prime}}} (2\delta p_{e^{\prime}}^{{\rm center}})
  +\frac{\partial M_{{\rm HYP}} }{\partial p_{K}} (2\delta p_{K}^{{\rm center}})\Bigr] \nonumber \\
  &=& -2 \delta p^{{\rm center}} \Bigl(
  \frac{\partial M_{{\rm HYP}} }{\partial p_{e}}
  +\frac{\partial M_{{\rm HYP}} }{\partial p_{e^{\prime}}} 
  +\frac{\partial M_{{\rm HYP}} }{\partial p_{K}} 
  \Bigr). \nonumber \\
\end{eqnarray}
The $\Delta M^{{\rm eloss}}_{{\rm HYP}}$ was calculated
taking into account the momentum loss 
for each particle, and found to be $\pm 0.31$, $\pm 0.20$, and $\pm 0.06$~MeV/$c^{2}$ for
the polyethylene, $^{7}$Li and $^{12}$C targets, respectively.


\subsection{Production point displacement from an assumed origin of BTM}
\label{sec:btm_displacement}
\begin{figure}[!htbp]
  \begin{center}
    \includegraphics[width=8.5cm]{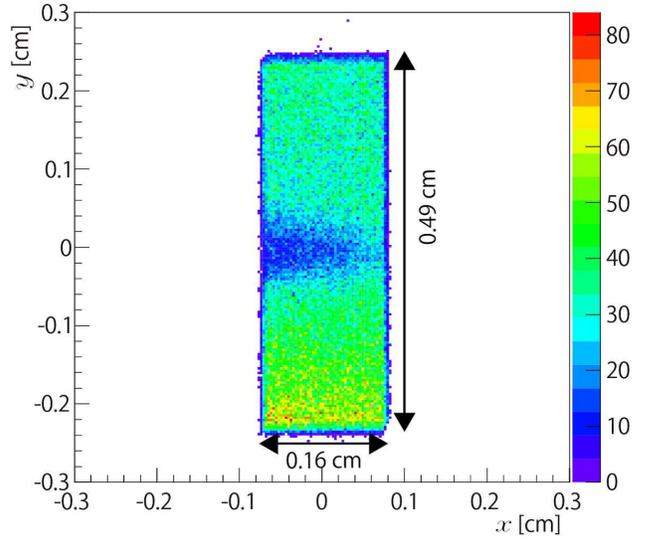}
    \caption{Beam profile on the polyethylene target.
      The beam was rastered in the area of $0.16^{x}\times0.46^{y}$~cm$^{2}$
      for the polyethylene target.}
    \label{fig:raster_ch2}
  \end{center}
\end{figure}

\begin{figure}[!htbp]
  \begin{center}
    \includegraphics[width=8.5cm]{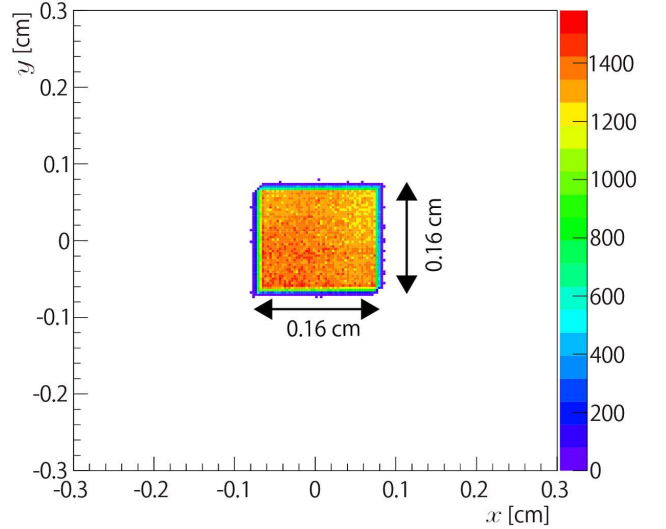}
    \caption{Beam profile on the $^{7}$Li target.
      The Beam was rastered in the area of $0.16^{x}\times0.16^{y}$~cm$^{2}$
      for the $^{7}$Li target.}
    \label{fig:raster_li7}
  \end{center}
\end{figure}

The BTMs were generated with an 
assumption that hypernuclei are produced at the point of the target center.
In the actual situation, however, 
the production points could be displaced from the assumed origin along with
the $z$-direction.
In addition, for the polyethylene and $^{7}$Li targets,
the beam rastering in the $x$ and $y$ directions
was applied in order to avoid melting of the targets from beam heating.
Figures~\ref{fig:raster_ch2} and \ref{fig:raster_li7} show the 
raster patterns for the polyethylene and $^{7}$Li targets for the actual data.
The raster patterns were obtained by measuring on an event by event basis
the voltages applied to the dipole magnets used for rastering.
The displacement of the production point from the assumed BTM origin
affects the missing mass resolution. 
It is noted that a counting rate 
around $0.05<y<0.05$~cm for the polyethylene target 
is low because the target was cracked due to heat
despite the rastering.
During the experiment,
the polyethylene target was moved to new position
every a few hours.  The raster pattern and trigger rates were monitored
in order to avoid serious damage on the target.


\subsubsection{$z$-dependence}
The target has a finite thickness, and thus, 
points where $\Lambda$ hypernuclei or hyperons are produced are varied 
event by event in the target along with the beam direction ($z$ direction).
We performed a Monte Carlo simulation 
to study an effect on the missing mass due to the $z$ displacement from the BTM origin. 
In the simulation, no target was placed and particles ($e$, $e^{\prime}$ and $K^{+}$) 
were randomly generated in the range of actual target thickness along with the $z$-direction.
As a result, the $z$ displacement was found to yield the missing-mass shift.
The missing-mass broadening $\Delta M^{{\rm Matrix}(z)}_{{\rm offset}}$ due to the mass shifts 
for the production of $\Lambda$, $^{7}_{\Lambda}$He and $^{12}_{\Lambda}$B
were found to be $\pm 0.37$, $\pm 0.34$, and $\pm 0.09$~MeV/$c^{2}$, respectively. 
\subsubsection{$x$ and $y$-dependence}
The beam was rastered for the polyethylene and $^{7}$Li targets because
their melting points were lower than the other targets.
The raster areas were $0.16^{x}\times0.46^{y}$~cm$^{2}$ and 
$0.16^{x}\times0.16^{y}$~cm$^{2}$ for these two targets.
(See Figs.~\ref{fig:raster_ch2} and \ref{fig:raster_li7})
Therefore, production points can be displaced from the BTM origin in the $x$ and 
$y$ directions for these targets. 
To investigate effects on the missing mass due to the displacements in $x$ and $y$ directions, 
Monte Carlo simulations were performed as was done for the $z$ displacement. 
The mass broadening due to the $x$ and $y$ displacements 
was found to be less than a few hundred keV. 
However, this effect can be removed because 
we measured a correlation of $x$ and $y$ positions versus the missing mass for $\Lambda$ 
in the data analysis.
The obtained correlation was used for corrections of the $x$ and $y$ displacements 
for the production of $\Lambda$ ($\Sigma^{0}$) and $^{7}_{\Lambda}$He.

\subsection{Comparison between the full estimation and obtained results}
\label{sec:mm_comp_simres}
\begin{figure}[!htbp]
  \begin{center}
    \includegraphics[width=8.6cm]{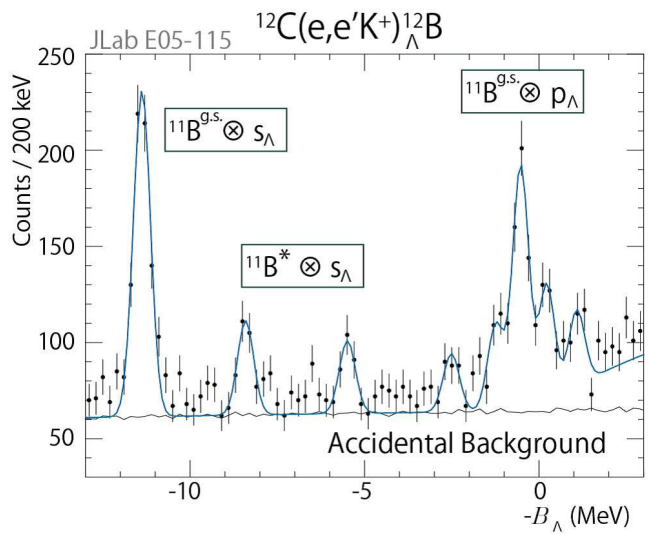}
    \caption{A binding energy  spectrum for the $^{12}$C{\eek}$^{12}_{\Lambda}$B reaction
      using a natural carbon target with a thickness of 0.088 g/cm$^{2}$ in JLab E05-115.
      Peak structures which correspond to
      ($^{11}$B$^{g.s.}\otimes s_{\Lambda}$),
      ($^{11}$B$^{*}\otimes s_{\Lambda}$),
      and ($^{11}$B$^{g.s.}\otimes p_{\Lambda}$) clearly can be seen with a FWHM of 0.54~MeV
      (refer to Ref.~\cite{cite:12LB}).
      }
    \label{fig:BL_12LB}
  \end{center}
\end{figure}

The missing-mass resolution cannot be simply estimated by
each contribution from the above sources because
some of them are not independent from each other.
In addition, the missing-mass resolution depends on 
achieved momentum and angular resolutions after the BTM optimization
(energy scale calibration, Sec.~\ref{sec:energy_calib}). 
Therefore, we performed a full modeled Monte Carlo simulation
to estimate the realistic missing-mass resolution.
In the simulation, the calibration analyses that were used for the real data 
were applied to various sets of dummy data and distorted BTMs
as described in Sec.~\ref{sec:energy_calib}.
Typical results obtained in the simulation and
the results of the real data analyses~\cite{cite:7LHe_2,cite:12LB}
are tabulated in Table~\ref{tab:reso_summary}, and 
these are fairly consistent.
Figure~\ref{fig:BL_12LB} shows
the obtained $B_{\Lambda}$ spectrum
for $^{12}_{\Lambda}$B~\cite{cite:12LB} with the energy
resolution of 0.54-MeV FWHM. 
\begin{table}[!htbp]
  \begin{center}
    \caption{A comparison of missing mass resolution
      between the Monte Carlo simulation and real data analyses
      for production of $\Lambda$, $^{7}_{\Lambda}$He, and $^{12}_{\Lambda}$B
      in JLab E05-115.}
    \label{tab:reso_summary}
    \begin{tabular}{l|l|ccc}
      \hline \hline
      \multicolumn{2}{l|}{Hyperon/Hypernucleus} & $\Lambda$ & $^{7}_{\Lambda}$He & $^{12}_{\Lambda}$B \\  \hline
      \multicolumn{2}{l|}{Target} & CH$_{2}$ & $^{7}$Li & $^{12}$C \\
      \multicolumn{2}{l|}{Thickness $($g/cm$^{2})$} & 0.451 & 0.208 & 0.088  \\ 
      \multicolumn{2}{l|}{Length in $z$ $($mm$)$} & 5.0 & 3.9 & 0.5 \\ \hline 
      \multicolumn{2}{l|}{$\Delta M^{{\rm int}}_{{\rm HYP}}$ FWHM (MeV/$c^{2}$)} & 0.73 & 0.41 & 0.41  \\ 
      \multicolumn{2}{l|}{$\Delta M^{{\rm Matrix}(z)}_{{\rm offset}}$ (MeV/$c^{2}$)} & 
      $\pm 0.37$ & $\pm0.34$ & $\pm 0.09$ \\ 
      \multicolumn{2}{l|}{$\Delta M^{{\rm eloss}}_{{\rm offset}}$ (MeV/$c^{2}$)}& 
      $\pm 0.31$ & $\pm 0.20$ & $\pm 0.06$ \\  \hline
      $\Delta M$  & Simulation & 1.6 & 1.3 &  0.5\\ \cline{2-5}
      FWHM  & Real data & 1.5 
      & 1.3~\cite{cite:7LHe_2} &  0.54~\cite{cite:12LB} \\
      (MeV/$c^{2}$)&  & (Fig.~\ref{fig:lambda_mm}) &  & (Fig.~\ref{fig:BL_12LB})\\
      \hline \hline
    \end{tabular}
  \end{center}
\end{table}


\section{Conclusion}
\label{sec:conclusion}
We study one of the fundamental forces, 
the strong force, by investigating
the $\Lambda N$ interaction through
spectroscopy of $\Lambda$ hypernuclei.
$\Lambda$-hypernuclear spectroscopy with the {\eek} reaction,
which complements experiments with other reactions, was
established in JLab. 
The unique features of the {\eek} experiment are
the higher energy resolution ($\Delta E \simeq 0.5$-MeV FWHM) and better accuracy of the
$\Lambda$ binding energy ($\Delta B_{\Lambda} \leq 0.2$~MeV)
compared to existing reaction spectroscopy with the {\kpi} and {\pik} reactions,
thanks to the primary electron beam at JLab.

A new spectrometer system consisting of
the SPL $+$ HES $+$ HKS was designed
to measure hypernuclei up to the medium heavy mass region ($A\leq52$) in
the latest hypernuclear experiment E05-115 at JLab Hall C.
In addition, we developed novel techniques
of semi-automated energy scale calibration
using $\Lambda$ and $\Sigma^{0}$ production from the hydrogen target.
The new spectrometer system and calibration technique
resulted in the best energy resolution and $B_{\Lambda}$ accuracy
(e.g. FWHM $=0.54$~MeV and $\Delta B^{sys.}_{\Lambda}=0.11$~MeV
for $^{12}$C{\eek}$^{12}_{\Lambda}$B) in reaction spectroscopy of $\Lambda$ hypernuclei,
and spectroscopic results for
$^{7}_{\Lambda}$He~\cite{cite:7LHe_2},
$^{10}_{\Lambda}$Be~\cite{cite:10LBe},
and $^{12}_{\Lambda}$B~\cite{cite:12LB} have been successfully obtained and published.
For the $^{52}$Cr target, on the other hand, $e^{-}e^{+}$ background events which
increased in proportion to a square of target proton number $Z^{2}$
caused high rates and high multiplicity in the HKS detector system.
The analysis for the $^{52}$Cr{\eek}$^{52}_{\Lambda}$V is in progress
under such a severe multiplicity condition.

The established techniques particularly energy calibration method
will be a basis and be further developed in the next hypernuclear experiments
at JLab~\cite{cite:pac44, cite:pac45, cite:loi208}. 



\section*{Acknowledgments}
At first, we thank the JLab staffs of the physics, accelerator,
and engineering divisions for support of the experiment.
We also express our appreciation to all of members
of HNSS (E89-009) and HKS (E01-011 and E05-115) Collaborations.
We thank M.~Sumihama and T.~Miyoshi for their contributions
to the early stage of the hypernuclear experiments at JLab Hall C.
We thank teams led by P.~Brindza and W.~Kellner
respectively for their large contributions
to designs of detector frames and vacuum systems, and
installation of the experimental equipment in JLab Hall C.
We appreciate efforts on design and fabrication
of the target system by I.~Sugai, D.~Meekins, and A.~Shichijo.
We thank N.~Chiga for his technical supports
at Tohoku University.
We thank Research Center for Electron Photon Science, Tohoku University (ELPH),
Cyclotron and Radioisotope Center, Tohoku University (CYRIC),
and KEK-PS for giving us opportunities of performance tests
for the detector and target systems.
The HKS experimental program at JLab was supported by JSPS KAKENHI Grants
No.~17H01121, No.~12002001, No.~15684005, No.~16GS0201,
Grant-in-Aid for JSPS fellow No.~24$\cdot$4123,
JSPS Core-to-Core Program No.~21002, and
JSPS Strategic Young Researcher Overseas Visits
Program for Accelerating Brain Circulation No.~R2201.
This program was partially supported by JSPS and DAAD under the
Japan-Germany Research Cooperative Program.
This work was supported by U.S. Department of Energy Contracts
No.~DE-AC05-84ER40150, No.~DE-AC05-06OR23177, No.~DE-FG02-99ER41065,
No.~DE-FG02-97ER41047, No.~DE-AC02-06CH11357, No.~DE-FG02-00ER41110, and
No.~DE-AC02-98CH10886, and U.S.-NSF Contracts No.~013815 and No.~0758095.


\begin{thebibliography}{00}

\bibitem{cite:hashimototamura} O.~Hashimoto and H.~Tamura, {\it Progress in Particle and Nuclear Physics} {\bf 57}, 564653 (2006).
\bibitem{cite:feliciello_nagae} A.~Feliciello and T.~Nagae, {\it Reports on Progress in Physics {\bf 78}}, 096301 (2015).
\bibitem{cite:gal_hun_mil} A.~Gal, E.V.~Hungerford, and D.J.~Millener, {\it Rev. Mod. Phys. {\bf 88}}, 035004 (2016).
\bibitem{cite:sugimura}H.~Sugimura {\it et al.} (J-PARC E10 Collaboration), {\it Physics Letters B {\bf 729}}, 39--44, (2014).
\bibitem{cite:yamasan}T.O.~Yamamoto {\it et al.} (J-PARC E13 Collaboration), 
  {\it Phys. Rev. Lett. {\bf 115}}, 222501 (2015).
\bibitem{cite:honda}R.~Honda {\it et al.} (J-PARC E10 Collaboration), {\it Phys. Rev. C {\bf 96}}, 014005 (2017).
\bibitem{cite:hyphi}T.R.~Saito {\it et al.}, {\it Nucl. Phys. A {\bf 881}}, 218--227 (2012).
\bibitem{cite:rappold}C.~Rappold {\it et al.} (HypHI Collaboration), {\it Phys. Rev. C {\bf 88}}, 041001(R) (2013).
\bibitem{cite:rappold2}C.~Rappold {\it et al.}, {\it Nucl. Phys. A {\bf 913}}, 170--184 (2013);
  {\it Phys. Lett. B {\bf 747}}, 129--134 (2015).
\bibitem{cite:star}Y.~Zhu (STAR Collaboration), {\it Nucl. Phys. A {\bf 904--905}}, 551c--554c (2013).
\bibitem{cite:alice}J.~Adam {\it et al.} (ALICE Collaboration), 
{\it Phys. Lett. B {\bf 754}}, 360 (2016).
\bibitem{cite:patrick}A.~Esser, S.~Nagao, F.~Schulz, P.~Achenbach {\it et al.} (A1 Collaboration),
{\it Phys. Rev. Lett. {\bf 114}}, 232501 (2015).
\bibitem{cite:florian}F.~Schulz, P.~Achenbach {\it et al.} (A1 Collaboration), 
{\it Nucl. Phys. A {\bf 954}}, 149 (2016).
\bibitem{cite:patrick_hyp}P.~Achenbach {\it et al.}, {\it JPS Conf. Proc. {\bf 17}}, 011001 (2017).   
\bibitem{cite:7LHe} S.N.~Nakamura, A.~Matsumura. Y.~Okayasu, T.~Seva, V.~M.~Rodriguez, P.~Baturin, L.~Yuan {\it et al.}
(HKS (JLab E01-011) Collaboration), {\it Phys. Rev. Lett. {\bf 110}}, 012502 (2013).
\bibitem{cite:12LB} L.~Tang, C.~Chen, T.~Gogami, D.~Kawama, Y.~Han {\it et al.} 
(HKS (JLab E05-115 and E01-011) Collaborations), {\it Phys. Rev. C {\bf 90}}, 034320 (2014).
\bibitem{cite:10LBe} T.~Gogami, C.~Chen, D.~Kawama {\it et al.} (HKS (JLab E05-115) Collaboration),
{\it Phys. Rev. C {\bf 93}}, 034314 (2016).
\bibitem{cite:7LHe_2} T.~Gogami, C.~Chen, D.~Kawama {\it et al.} (HKS (JLab E05-115) Collaboration), 
{\it Phys. Rev. C {\bf 94}}, 021302(R) (2016).

\bibitem{cite:danysz_p} M.~Danysz and J.~Pniewski, {\it Phil. Mag. {\bf 44}} (1953).
\bibitem{cite:juric} M.~Juri$\check{{\rm c}}$ \textit{et al.}, {\it Nucl. Phys. B {\bf 52}}, 1 (1973).
\bibitem{cite:hasegawa2} T.~Hasegawa \textit{et al.}, \textit{Phys. Rev. C} \textbf{53}, 1210 (1996).
\bibitem{cite:hotchi} H.~Hotchi {\it et al.}, {\it Phys. Rev. C {\bf 64}}, 044302 (2001).
\bibitem{cite:12LC_1} P.~Dluzewski {\it et al.}, {\it Nucl. Phys. A {\bf 484}}, 520 (1988).
\bibitem{cite:12LC_2} D.H.~Davis, {\it Nucl. Phys. A {\bf 754}}, 3c--13c (2005).
\bibitem{cite:finuda} E.~Botta, T.~Bressani, A.~Feliciello, {\it Nucl. Phys. A {\bf 960}}, 165--179 (2017).
\bibitem{cite:mot_sot_ito} T.~Motoba, M.~Sotona, and K.~Itonaga, 
{\it Prog. Theor. Phys. Suppl. {\bf 117}}, 123 (1994).
\bibitem{cite:tamura}H.~Tamura {\it et al.}, {\it Phys. Rev. Lett. {\bf 84}}, 5963 (2000).
\bibitem{cite:tanida}K.~Tanida {\it et al.}, {\it Phys. Rev. Lett. {\bf 86}}, 1982 (2001).
\bibitem{cite:akikawa}H.~Akikawa {\it et al.}, {\it Phys. Rev. Lett. {\bf 88}}, 082501 (2002).
\bibitem{cite:ukai}M.~Ukai {\it et al.}, {\it Phys. Rev. Lett. {\bf 93}}, 232501 (2004);
{\it Phys. Rev. C {\bf 73}}, 012501(R) (2006); {\it Phys. Rev. C {\bf 77}}, 054315 (2008).
\bibitem{cite:hosomi}K.~Hosomi {\it et al.}, {\it Prog. Theor. Exp. Phys.}, 081D01 (2015).
\bibitem{cite:yang_tamura}S.B.~Yang {\it et al.}, {\it JPS Conf. Proc. {\bf 17}}, 012004 (2017);
H.~Tamura {\it et al.}, {\it JPS Conf. Proc. {\bf 17}}, 011004 (2017).
\bibitem{cite:pdg} K.A.~Olive {\it et al.} (Particle Data Group), {\it Chin. Phys. C {\bf 38}}, 090001 (2014) and 2015 update.

\bibitem{cite:gal} A.~Gal, \textit{Phys. Lett. B {\bf 744}}, 352--357 (2015).
\bibitem{cite:dani} D.~Gazda and A.~Gal, {\it Phys. Rev. Lett. {\bf 116}}, 122501 (2016); {\it Nucl. Phys. A {\bf 954}}, 161 (2016).
\bibitem{cite:yamamoto_neustar} Y.~Yamamoto, T.~Furumoto, N.~Yasutake, and Th.A.~Rijken, 
{\it Phys. Rev. C {\bf 90}}, 045805 (2014).
\bibitem{cite:isaka_def} M.~Isaka, K.~Fukukawa, M.~Kimura, E.~Hiyama, H.~Sagawa, and Y.~Yamamoto, 
{\it Phys. Rev. C {\bf 89}}, 024310 (2014).
\bibitem{cite:win} M.T.~Win and K.~Hagino, {\it Phys. Rev. C {\bf 78}}, 054311 (2008).
\bibitem{cite:bnlu} B.N.~Lu, E.~Hiyama, H.~Sagawa, and S.~G.~Zhou,
{\it Phys. Rev. C {\bf 89}}, 044307 (2014).
\bibitem{cite:xue} W.X.~Xue, J.M.~Yao, K.~Hagino, Z.P.~Li, H.~Mei, and Y.~Tanimura, 
{\it Phys. Rev. C {\bf 91}}, 024327 (2015).
\bibitem{cite:hiyama1} E.~Hiyama, Y.~Yamamoto, T.~Motoba, M.~Kamimura, \textit{Phys. Rev. C} \textbf{ 80}, 054321 (2009).
\bibitem{cite:hiyama_10LB} E.~Hiyama and Y.~Yamamoto, \textit{Prog. Theor. Phys.} \textbf{128}, 1 (2012).  
\bibitem{cite:bydzovsky} P.~Byd${\rm \check{z}}$ovsk${\rm \acute{y}}$,
M.~Sotona, T.~Motoba, K.~Itonaga, K.~Ogawa, and O.~Hashimoto, 
{\it Nucl. Phys. A {\bf 881}}, 199 (2012).
\bibitem{cite:kusaka-fujita}
{\it Master's Theses}, Tohoku University, Sendai, Japan (all in Japanese):
J.~Kusaka, ``{\it Design of an experiment for electro-production spectroscopy of medium-heavy hypernuclei}'', 2014;
M.~Fujita, ``{\it Design of an experiment for Lambda hypernuclear reaction
spectroscopy in wide mass range by the {\eek} reaction}'', 2016.
\bibitem{cite:loi208} O.~Benhar, F.~Garibaldi, T.~Gogami, E.C.~Markowitz, S.N.~Nakamura, J.~Reinhold,
L.~Tang, G.M.~Urciuoli, I.~Vidana,
{\it Letter of Intend to JLab PAC45}, LOI12-17-003,
``{\it Studying $\Lambda$ interactions in nuclear matter with the $^{208}$Pb{\eek}$^{208}_{\Lambda}$Tl reaction}'', 2017.
\bibitem{cite:toshi}T.~Gogami, {\it Doctoral Thesis},
``{\it Spectroscopic research of  hypernuclei up to medium-heavy mass region with the {\eek} reaction}'',
Tohoku University, Sendai, Japan, 2014.
\bibitem{cite:miyoshi} T.~Miyoshi {\it et al.} (HNSS Collaboration), {\it Phys. Rev. Lett. {\bf 90}}, 232502 (2003).
\bibitem{cite:lulin} L.~Yuan {\it et al.} (HNSS Collaboration), {\it Phys. Rev. C {\bf 73}}, 044607 (2006).
\bibitem{cite:28LAl} O.~Hashimoto {\it et al.}, {\it Nucl. Phys. A {\bf 835}}, 121--128 (2010).
\bibitem{cite:52LV} S.N.~Nakamura, T.~Gogami, and L.~Tang, {\it JPS Conf. Proc. {\bf 17}}, 011002 (2017). 
\bibitem{cite:9LLi} G.M.~Urciuoli {\it et al.} (Jefferson Lab Hall A Collaboration), 
{\it Phys. Rev. C {\bf 91}}, 034308 (2015).
\bibitem{cite:iodice} M.~Iodice {\it et al.} (Jefferson Lab Hall A Collaboration), {\it Phys. Rev. Lett. {\bf 99}}, 052501 (2007).
\bibitem{cite:cusanno} F.~Cusanno {\it et al.} (Jefferson Lab Hall A Collaboration), {\it Phys. Rev. Lett. {\bf 103}}, 202501 (2009).





\bibitem{cite:sotona} M.~Sotona and S.~Frullani, \textit{Prog. Theor. Phys. Suppl. {\bf 177}}, 151 (1994).
\bibitem{cite:hunger} E.V.~Hungerford, \textit{Prog. Theor. Phys. Suppl. {\bf 117}}, 135 (1994).
\bibitem{cite:xu} G.~Xu and E.V.~Hungerford, \textit{Nucl. Instrum. Methods Phys. Res. Sect. A {\bf 501}}, 602--614 (2003).
\bibitem{cite:bradford} R.~Bradford \textit{et al.}, \textit{Phys. Rev. C {\bf 73}}, 035202 (2006).
\bibitem{cite:gogami} T.~Gogami {\it et al.}, {\it Nucl. Instrum. Methods Phys. Res. Sect. A {\bf 729}}, 816--824 (2013).
\bibitem{cite:fujii} Y.~Fujii {\it et al.},  {\it Nucl. Instrum. Methods Phys. Res. Sect. A {\bf 795}}, 351--363  (2015).
\bibitem{cite:tsai} Y.~Tsai, \textit{Rev. Mod. Phys.} \textbf{46}, 4 (1974).
\bibitem{cite:moller} C.~Itzykson, J.~Zuber, \textit{Quantum Field Theory} (Dover, 1980).  
\bibitem{cite:miyoshi_dthesis} T.~Miyoshi, \textit{Doctoral Thesis},
``{\it Spectroscopic study of $\Lambda$ hypernuclei by the {\eek} reaction}'',
Tohoku University, Sendai, Japan, 2002.
\bibitem{cite:tosca}COBHAM [http://www.cobham.com].
\bibitem{cite:havar}Goodfellow [http://www.goodfellow.com].
\bibitem{cite:shichijo} A.~Shichijo, \textit{Master's Thesis},
``{\it Development of the target system and analysis of the $p${\eek}$\Lambda,\Sigma^{0}$ reactions in JLab E05-115 experiment}'',
Tohoku University, Sendai, Japan, 2010 (in Japanese).
\bibitem{cite:ansys}  ANSYS [http://ansys.jp].
\bibitem{cite:rika} Chronological Scientific Tables (2017) [http://www.rikanenpyo.jp].
\bibitem{cite:yokota}K.~Yokota, \textit{Master's Thesis},
``{\it Studies of scattered electron spectrometer and trigger logic in the third generation hypernuclear experiment at JLab Hall C}'',
Tohoku University, Sendai, Japan, 2009 (in Japanese).
\bibitem{cite:altera} ALTERA [http://www.altera.com].



\bibitem{cite:taniya} N.~Taniya, {\it Master's Thesis}, ``{\it Study of Cherenkov Counter for JLab E05-115 experiment}'',
Tohoku University, Sendai, Japan, 2010 (in Japanese).
\bibitem{cite:okayasu_d} Y.~Okayasu, {\it Doctoral Thesis}, ``{\it Spectroscopic study of light Lambda hypernuclei via the {\eek} reaction}'',
Tohoku University, Sendai, Japan, 2008.
\bibitem{cite:hrs}J.~Alcorn {\it et al.}, {\it Nucl. Instrum. Methods Phys. Res. Sect. A {\bf 522}}, 294--346 (2004).
\bibitem{cite:pac44}F.~Garibaldi, P.E.C.~Markowitz,
S.N.~Nakamura, J.~Reinhold, L.~Tang, G.M.~Urciuoli (spokespersons) (JLab Hypernuclear Collaboration),
{\it Proposal to JLab PAC44}, E12-15-008,
``{\it An isospin dependence study of the $\Lambda$-N interaction through the high precision spectroscopy of $\Lambda$-hypernuclei with electron beam}'',  
2016.  



\bibitem{cite:pac45}L.~Tang, F.~Garibaldi, P.E.C.~Markowitz,
S.N.~Nakamura, J.~Reinhold, G.M.~Urciuoli (spokespersons)
(JLab Hypernuclear Collaboration),
{\it Proposal to JLab PAC45}, E12-17-003,
``{\it Determining the Unknown $\Lambda$-n Interaction by Investigating the $\Lambda$nn Resonance}'', 2017.


\end{thebibliography}
\end{document}